\documentclass[11pt]{article}

\usepackage{epsf}
\usepackage{epsfig}
\usepackage{graphicx}
\usepackage{rotating}
\usepackage{refmerge} 
\usepackage{supertabular}
\usepackage{subfigure}
\usepackage{multirow}
\usepackage{refmerge} 
\usepackage{lscape}

\newcommand{\BABARPubYear}    {04}

\newcommand{\BABARConfNumber} {38}
\newcommand{\SLACPubNumber} {10658}

\input pubboard/babarsym

\newcommand{\bei}{\begin{itemize}}
\newcommand{\eei}{\end{itemize}}
\newcommand{\beq}{\begin{equation}}
\newcommand{\eeq}{\end{equation}}
\newcommand{\beqn}{\begin{eqnarray}}
\newcommand{\eeqn}{\end{eqnarray}}
\newcommand{\beqns}{\begin{eqnarray*}}
\newcommand{\eeqns}{\end{eqnarray*}}

\newcommand{\intl}{\int\limits}

\newcommand{\e}{\epsilon}

\def\bk{\!\!\!\!}

\def\PRL{{\em Phys. Rev. Lett.}}

\def\ea{{\em et al.}}
\def\min{{\rm min}}

\def\rPTbarkappa{\kern 0.18em\overline{\kern -0.18em r}{}^{\kappa}{}}
\def\rPTbarsigma{\kern 0.18em\overline{\kern -0.18em r}{}^{\sigma}{}}

\def\deltabarkappa{\kern 0.18em\overline{\kern -0.18em \delta}{}_r^{\kappa}}
\def\deltabarsigma{\kern 0.18em\overline{\kern -0.18em \delta}{}_r^{\sigma}}
\def\deltaTbarkappa{\kern 0.18em\overline{\kern -0.18em \delta}{}_T^{\kappa}}
\def\deltaTbarsigma{\kern 0.18em\overline{\kern -0.18em \delta}{}_T^{\sigma}}

\newcommand\ph{\phantom}
\newcommand{\half}{\ensuremath{{1\over2}}}
\newcommand{\pvec}{{\bf p}}

\def\OC{X}
\def\OCbar{{\kern 0.18em\overline{\kern -0.18em \OC}}}
\def\mBz{m_{\Bz}}
\def\spz{s_{+}}
\def\smz{s_{-}}
\def\spm{s_{0}}

\def\mpm{m_{0}}

\def\fpz{f_{+}}
\def\fmz{f_{-}}
\def\fpm{f_{0}}

\def\mpmMax{\mpm^{\rm max}}
\def\mpmMin{\mpm^{\rm min}}

\def\pipipi{\pip\pim\piz}
\def\Btopipipi{\Bz\to\pipipi}

\def\mprime{m^\prime}
\def\thetaprime{\theta^\prime}
\def\deprime{{\de^\prime}{}}
\def\dt{\deltat}

\def\de{\DeltaE}
\def\demin{\de_{-}}
\def\demax{\de_{+}}
\def\deminmax{\de_{\pm}}

\def\tpi{3\pi}

\def\Qtag{q_{\rm tag}}
\def\Qtagi{q_{{\rm tag},i}}
\def\Atagqq{A_{q\bar q,\,\rm tag}}
\def\Atag{A_{{B^+,\,\rm tag}}}
\def\Atagj{A_{{B^+,\,\rm tag},j}}

\def\cat{c}

\def\detJ{|\det J|}
\def\detJi{|\det J_i|}
\def\Rscf{R_{\rm SCF}}

\def\ampfrac{a}
\renewcommand\a{\kappa}

\newcommand\cch{\tau}

\def\Amptp{{\cal A}_{3\pi}}
\def\Amptpbar{\kern 0.18em\overline{\kern -0.18em {\cal A}}_{3\pi}}

\def\absAmptp{|\Amptp|}
\def\absAmptpbar{|\Amptpbar|}
\def\Amptpkappa{{A^{\kappa}}}
\def\Amptpsigma{{A^{\sigma}}}
\def\Amptpbarkappa{\kern 0.18em\overline{\kern -0.18em A}{}^{\kappa}{}}
\def\Amptpbarsigma{\kern 0.18em\overline{\kern -0.18em A}{}^{\sigma}{}}
\def\AmpAll{|{\cal A}_{3\pi}^\pm(\dmt)|^2}
\def\AmpAllp{|{\cal A}_{3\pi}^+(\dmt)|^2}
\def\AmpAllm{|{\cal A}_{3\pi}^-(\dmt)|^2}

\def\Tbarkappa{\kern 0.18em\overline{\kern -0.18em T}{}^{\kappa}{}}
\def\Tbarsigma{\kern 0.18em\overline{\kern -0.18em T}{}^{\sigma}{}}
\def\Pbarkappa{\kern 0.18em\overline{\kern -0.18em P}{}^{\kappa}{}}
\def\Pbarsigma{\kern 0.18em\overline{\kern -0.18em P}{}^{\sigma}{}}
\def\kappab{\overline\kappa}

\def\jchich{\kappab}
\def\ichjch{\kappa}
\newcommand\Aij{{A^{\ichjch}}}
\def\Abij{\kern 0.18em\overline{\kern -0.18em A}{}^{\ichjch}{}}
\def\Tij{{T^{\ichjch}}}
\def\Tji{{T^{\jchich}}}
\def\Pij{{P^{\ichjch}}}
\def\Pji{{P^{\jchich}}}
\def\R{{\rm Re}}
\def\I{{\rm Im}}
\def\kappm{\kappa^{+-}}
\def\kapmp{\kappa^{-+}}

\def\bbar{\Bz \Bzb}

\def\Crhopi{C}
\def\dCrhopi{\Delta C}

\def\dS{\Delta S}
\def\dC{\Delta C}
\def\rhopi{\rho\pi}
\def\Acp{{\cal A}_{\rho\pi}}

\def\Acppm{{\cal A}_{\rho\pi}^{+-}}
\def\Acpmp{{\cal A}_{\rho\pi}^{-+}}

\def\Nbpm{{\kern 0.18em\overline{\kern -0.18em N}}^{+-}}
\def\Nbmp{{\kern 0.18em\overline{\kern -0.18em N}}^{-+}}
\def\rar{\rightarrow}

\def\Mu{\mu}
\def\Chi2MinaMu{\chi^2_{\min ;\a,\Mu}}
\def\Chi2MinMu{\chi^2_{\min ;\Mu}(a)}

\def\dmt{\Delta t}
\def\dmd{\Delta m_d}

\def\TM{{\rm TM}}
\def\SCF{{\rm SCF}}

\def\fscfave{\kern 0.18em\overline{\kern -0.18em f}_{\rm SCF}}
\def\fscf{f_{\rm SCF}}
\def\fscfi{f_{{\rm SCF},i}}

\def\abar{\bar{a}}

\def\PRL{{\em Phys. Rev. Lett.}}

\def\Bbar{\kern 0.18em\overline{\kern -0.18em B}{}\xspace}

\def\BRpmb{{\cal \kern 0.18em\overline{\kern -0.18em  B}}{}_{\rho\pi}^{+-}}
\def\BRmpb{{\cal \kern 0.18em\overline{\kern -0.18em  B}}{}_{\rho\pi}^{-+}}

\def\BRipmb{{\cal \kern 0.18em\overline{\kern -0.18em  B}}{}_{\rho^+\pi^-}}
\def\BRimpb{{\cal \kern 0.18em\overline{\kern -0.18em  B}}{}_{\rho^-\pi^+}}

\def\Abar{\kern 0.18em\overline{\kern -0.18em A}{}}
\def\abar{\kern 0.18em\overline{\kern -0.18em a}{}}

\def\Apm{A^{+}}
\def\Amp{A^{-}}
\def\Apmb{\Abar^{+}}
\def\Ampb{\Abar^{-}}

\def\Azz{A^{0}}
\def\Azzb{\Abar^{0}}

\def\ie{{\em i.e.}} 
\def\cf{{\em cf.}} 
\def\eg{{\em e.g.}}

\newcommand{\UPm    }{\ensuremath{U_+^-}}
\newcommand{\UMp    }{\ensuremath{U_-^+}}
\newcommand{\UMm    }{\ensuremath{U_-^-}}
\newcommand{\UPMpRe }{\ensuremath{U_{+-}^{+,\R}}}
\newcommand{\UPMmRe }{\ensuremath{U_{+-}^{-,\R}}}
\newcommand{\UPMpIm }{\ensuremath{U_{+-}^{+,\I}}}
\newcommand{\UPMmIm }{\ensuremath{U_{+-}^{-,\I}}}
\newcommand{\Uzp    }{\ensuremath{U_0^+}}
\newcommand{\UPzpRe }{\ensuremath{U_{+0}^{+,\R}}}
\newcommand{\UPzpIm }{\ensuremath{U_{+0}^{+,\I}}}
\newcommand{\UMzpRe }{\ensuremath{U_{-0}^{+,\R}}}
\newcommand{\UMzpIm }{\ensuremath{U_{-0}^{+,\I}}}
\newcommand{\IP     }{\ensuremath{I_+}}
\newcommand{\IM     }{\ensuremath{I_-}}
\newcommand{\IPMRe  }{\ensuremath{I_{+-}^{\R}}}
\newcommand{\IPMIm  }{\ensuremath{I_{+-}^{\I}}}

\long\def\inst#1{\par\nobreak\kern 4pt\nobreak
    {\it #1}\par\vskip 10pt plus 3pt minus 3pt}

\begin{document}

{\pagestyle{empty}

\begin{flushright}
\babar-CONF-\BABARPubYear/\BABARConfNumber \\
SLAC-PUB-\SLACPubNumber \\
August 2004 \\
\end{flushright}

\par\vskip 0.5cm

\begin{center}
\Large \bf 
Measurement of {\em CP}-Violating Asymmetries in 
\boldmath$B^0\to(\rho\pi)^0$ \\[0.05cm]
Using a Time-Dependent Dalitz Plot Analysis
\end{center}
\bigskip

\begin{center}
\large The \babar\ Collaboration\\
\mbox{ }\\
\today
\end{center}
\bigskip \bigskip

\begin{center}
\large \bf Abstract
\end{center}
We present the preliminary measurement of \CP-violating asymmetries in 
$B^0\to(\rho\pi)^0\to\pi^+\pi^-\pi^0$ decays using a time-dependent 
Dalitz plot analysis. The results are obtained from a data sample 
of 213 million $\FourS \to B\Bbar$ decays, collected by the \babar\ 
detector at the \pep2\ asymmetric-energy \B~Factory at SLAC. This 
analysis extends the narrow-$\rho$ quasi-two-body approximation 
used in the previous analysis, by taking into account the 
interference between the $\rho$ resonances of the three charges. 
We measure 16 coefficients of the bilinear form factor terms occurring 
in the time-dependent decay rate of the \Bz meson with the use of a 
maximum-likelihood fit. We derive the physically relevant quantities 
from these coefficients. 
We measure the direct \CP-violation parameters
$\Acp=-0.088\pm 0.049\pm{0.013}$ and
$C=0.34\pm 0.11\pm{0.05}$, where the first errors are statistical
and the second systematic.
For the mixing-induced \CP-violation parameter we find
$S=-0.10\pm 0.14 \pm{0.04}$, and
for the dilution and strong phase shift parameters 
respectively, we obtain $\dC=0.15\pm 0.11 \pm{0.03}$
and $\dS=0.22\pm 0.15\pm{0.03}$. For the angle $\alpha$ 
of the Unitarity Triangle we measure $(113^{\,+27}_{\,-17}\pm6)^\circ$,
while only a weak constraint is achieved at the significance level of 
more than two standard deviations. 
Finally, for the relative strong phase $\delta_{+-}$ 
between the $\Bz\to\rho^-\pi^+$ and $\Bz\to\rho^+\pi^-$ transitions we find 
$(-67^{\,+28}_{\,-31}\pm7)^\circ$, with a similarly weak constraint
at two standard deviations and beyond.

\vfill
\begin{center}

Submitted to the 32$^{\rm nd}$ International Conference on High-Energy Physics, ICHEP 04,\\
16 August---22 August 2004, Beijing, China

\end{center}

\vspace{1.0cm}
\begin{center}
{\em Stanford Linear Accelerator Center, Stanford University, 
Stanford, CA 94309} \\ \vspace{0.1cm}\hrule\vspace{0.1cm}
Work supported in part by Department of Energy contract DE-AC03-76SF00515.
\end{center}

\newpage
} 

%
%

\begin{center}
\small

The \babar\ Collaboration,
\bigskip

%
B.~Aubert,
R.~Barate,
D.~Boutigny,
F.~Couderc,
J.-M.~Gaillard,
A.~Hicheur,
Y.~Karyotakis,
J.~P.~Lees,
V.~Tisserand,
A.~Zghiche
\inst{Laboratoire de Physique des Particules, F-74941 Annecy-le-Vieux, France }
% ...

\end{center}\newpage

\section{INTRODUCTION}
\label{sec:Introduction}

Measurements of the parameter $\stwob$~\cite{BaBarSin2beta,BelleSin2beta} 
have established \CP violation in the $\Bz$ meson system 
and provide strong support for the Kobayashi and Maskawa model of 
this phenomenon as arising from a single phase in the three-generation
CKM quark-mixing matrix~\cite{CKM}. 
We present in this letter preliminary results from a time-dependent 
analysis of the $\Bz\to\pip\pim\piz$ Dalitz plot (DP) that is dominated 
by the $\rho$ intermediate resonances. The goal of the analysis
is the simultaneous extraction of the strong transition 
amplitudes and the weak interaction phase 
$\alpha\equiv \arg\left[-V_{td}^{}V_{tb}^{*}/V_{ud}^{}V_{ub}^{*}\right]$
of the Unitarity Triangle. In the Standard Model, a non-zero 
value for $\alpha$ would be responsible for the occurrence 
of mixing-induced \CP violation in this decay.
The \babar\  and Belle experiments have obtained constraints
on $\alpha$ from the measurement of effective quantities $\stwoa_{\rm eff}$
in $B$ decays to $\pip\pim$~\cite{babarpipi,bellepipi} and \babar\
from $B$ decays to $\rho^+\rho^-$~\cite{babarrhorho}, using an 
isospin analysis~\cite{GLisospin} that involves the other charges
of these final states to obtain bounds on $\alpha-\alpha_{\rm eff}$.

Unlike $\pip\pim$, $\rho^{\pm}\pi^{\mp}$ is not a \CP 
eigenstate, and four flavor-charge configurations
$(\Bz(\Bzb) \to \rho^{\pm}\pi^{\mp})$ must be considered.  
The corresponding isospin analysis~\cite{Lipkinetal} is unfruitful
with the present statistics since two pentagonal amplitude relations with 
12 unknowns have to be solved (compared to 6 unknowns for
the  $\pip\pim$ and $\rho^+\rho^-$ systems). However, it
has been pointed out by Snyder and Quinn~\cite{SnyderQuinn}, that 
one can obtain the necessary degrees of freedom to constrain
$\alpha$ without ambiguity by explicitly including in the analysis the 
variation of the strong phases of the interfering $\rho$ resonances
in the Dalitz plot.

The present analysis focuses on the decay $B^0\to(\rho\pi)^0\to\pip\pim\piz$. 
The data sample used is a superset of that used for our previous 
time-dependent result~\cite{rhopipaper}, obtained within the 
quasi-two-body approximation. In that approach, the analysis was 
restricted to the charged-$\rho$ regions in the Dalitz plot, and the 
interference regions were removed. Here, we extend the analysis to 
the entire region of interest in the Dalitz plot, which contains
the $\rho$ resonances of all three charges and their interference.

\subsection{DECAY AMPLITUDES}
\label{sec:kinmeatics}

We consider the decay of a spin-zero $\Bz$ with four-momentum
$p_B$ into the three daughters $\pip(p_+)$, $\pim(p_-)$, $\piz(p_0)$,
with corresponding four-momenta. Using
as independent (Mandelstam) variables the invariant masses-squared
\beq
\label{eq:dalitzVariables}
       \spz \;=\; (p_+ + p_0)^2~, \hspace{1cm}
       \smz \;=\; (p_- + p_0)^2~, 
\eeq
the invariant mass of the positive and negative pion, 
$\spm \;=\; (p_+ + p_-)^2$, is obtained from energy and 
momentum conservation
\beq
\label{eq:magicSum}
	\spm \;=\; \mBz^2 + 2m_{\pi^+}^2 + m_{\pi^0}^2
		   - \spz - \smz~.
\eeq
The differential $\Bz$ decay width with respect to the 
variables defined in Eq.~(\ref{eq:dalitzVariables}) (\ie, the 
{\em Dalitz plot}) reads
\beq
\label{eq:partialWidth}
	d\Gamma(\Btopipipi) \;=\; 
	\frac{1}{(2\pi)^3}\frac{|\Amptp|^2}{8 \mBz^3}\,d\spz d\smz~,
\eeq
where $\Amptp$ is the Lorentz-invariant amplitude
of the three-body decay. 

We assume in the following that the amplitudes $\Amptp$ and its complex 
conjugate $\Amptpbar$, corresponding to the transitions $\Bz\to\pip\pim\piz$ 
and $\Bzb\to\pip\pim\piz$, respectively, are dominated by the three 
resonances $\rho^+$, $\rho^-$ and $\rho^0$. The $\rho$ resonances
are assumed to be the sum of the ground state $\rho(770)$ and the
radial excitations $\rho(1450)$ and $\rho(1700)$, with an initial set
of resonance parameters and relative amplitudes determined by a combined fit 
to $\tau^+\to\nutb\pip\piz$ and $\epem\to\pip\pim$ data~\cite{taueeref}.
Since the hadronic environment is different in \B decays, we 
cannot rely on this result and therefore determine the relative $\rho(1450)$
amplitude simultaneously with the \CP parameters from the fit. Variations of
the other parameters and possible contributions to the $\Bz\to\pip\pim\piz$ 
decay other than the $\rho$'s are studied as part of the systematic 
uncertainties (Section~\ref{sec:Systematics}).

Including the $\BzBzb$ mixing parameter $q/p$ into the $\Bzb$ decay 
amplitudes, we can write~\cite{SnyderQuinn,BaBarPhysBook}
\beqn
\label{eq:amp}
   \Amptp    		
	&=& \fpz \Apm + \fmz \Amp + \fpm\Azz ~, \\
\label{eq:ampBar}
   \Amptpbar 	
	&=& \fpz \Apmb + \fmz \Ampb + \fpm\Azzb ~,
\eeqn
where the $f_\kappa$ (with the $\rho$ charge $\kappa=\{+,-,0\}$) 
are functions of the Dalitz variables 
$\spz$ and $\smz$ that incorporate the kinematic and dynamical properties 
of the $\Bz$ decay into a (vector) $\rho$ resonance and a 
(pseudoscalar) pion, and where the $\Aij$ are 
complex amplitudes\footnote
{
	The amplitude superscript ``$\kappa$''
	denotes the charge of the $\rho$ from the decay 
	of the $\Bz$ meson.
}
that
may comprise weak and strong transition phases and that are independent 
of the Dalitz variables. Note that the definitions~(\ref{eq:amp})
and (\ref{eq:ampBar}) imply the assumption that the relative phases 
between the $\rho(770)$ and its radial excitations are \CP-conserving.

Following Ref.~\cite{taueeref}, the $\rho$
resonances are parameterized by a modified relativistic Breit-Wigner 
function introduced by Gounaris and Sakurai (GS)~\cite{rhoGS}. Due to angular 
momentum conservation, the spin-one $\rho$ resonance is polarized 
in a helicity-zero state. For a $\rho^{\a}$ resonance with charge $\a$,
the GS function is multiplied by the kinematic function 
$-4|{\bf p}_\a||{\bf p}_\cch|\cos\theta_{\a}$, where the momenta
are defined in the $\rho$-resonance rest frame, and where
${\bf p}_\cch$ is the momentum of the 
particle not from $\rho$ decay,
and $\cos\theta_{\a}$ the cosine of the helicity angle of
the $\rho^{\a}$. For the $\rho^+$ ($\rho^-$), $\theta_{+}$ 
($\theta_{-}$) is defined by the angle between the $\pi^0$ ($\pi^-$) 
in the $\rho^+$ ($\rho^-$) rest frame and the $\rho^+$ ($\rho^-$) 
flight direction in the $\Bz$ rest frame. For the $\rho^0$, 
$\theta_{0}$ is defined by the angle between the $\pi^+$ in 
the $\rho^0$ rest frame and the $\rho^0$ flight direction in 
the $\Bz$ rest frame. With these definitions, each pair of GS functions 
interferes destructively at equal masses-squared.

The occurrence of $\cos\theta_{\a}$ 
in the kinematic functions substantially enhances the interference 
between the different $\rho$ bands in the Dalitz plot, and thus 
increases the sensitivity of this analysis~\cite{SnyderQuinn}.

\subsection{TIME DEPENDENCE}

With $\deltat \equiv t_{\tpi} - t_{\rm tag}$ defined as the proper 
time interval between the decay of the fully reconstructed $B^0_{\tpi}$ 
and that of the  other meson $\Bz_{\rm tag}$,  the time-dependent decay
rate $\AmpAllp$ ($\AmpAllm$) when the tagging meson is a $\Bz$ ($\Bzb$) 
is given by 
\beqn
\label{eq:dt}
    \AmpAll
	&=& 
		\frac{e^{-|\dmt|/\tau_{B^0}}}{4\tau_{B^0}}
	\bigg[\absAmptp^2 + \absAmptpbar^2
	      \mp \left(\absAmptp^2 - \absAmptpbar^2\right)\cos(\dmd\dmt)	
	\nonumber\\
	&&\phantom{\frac{e^{-|\dmt|/\tau_{B^0}}}{4\tau_{B^0}}
	\bigg[\absAmptp^2 + \absAmptpbar^2}	
	      \pm\,2\I\left[\Amptpbar\Amptp^*\right]\sin(\dmd\dmt)	
	\bigg]~,
\eeqn
where $\tau_{B^0}$ is the mean \Bz lifetime, $\deltamd$ is the $\BzBzb$ 
oscillation frequency, and where we have assumed that \CP violation in $\bbar$ 
mixing is absent ($|q/p|=1$), $\Delta\Gamma_{B_d}=0$ and \CPT is conserved.
Inserting the amplitudes~(\ref{eq:amp}) and (\ref{eq:ampBar}), one 
obtains for the terms in Eq.~(\ref{eq:dt})
\beqn
   \label{eq:UI}
   \absAmptp^2 \pm \absAmptpbar^2 
	&=&
	\sum_{\kappa\in\{+,-,0\}}  |f_\kappa|^2U_\kappa^\pm
	\;\;+ \;\;
	2\bk\bk\sum_{\kappa <\sigma\in\{+,-,0\}} 
	\left(
	    \,\R\left[f_\kappa f_\sigma^*\right]U_{\kappa\sigma}^{\pm,\R}
	  - \,\I\left[f_\kappa f_\sigma^*\right]U_{\kappa\sigma}^{\pm,\I}
	\right)~,
	\nonumber\\[0.2cm]
   \I\left(\Amptpbar\Amptp^\star\right)
	&=&
	\sum_{\kappa\in\{+,-,0\}}  |f_\kappa|^2I_\kappa
	\;\;+ 
	\sum_{\kappa <\sigma\in\{+,-,0\}} 
	\left(
	    \,\R\left[f_\kappa f_\sigma^*\right]I_{\kappa\sigma}^{\I}
	  + \,\I\left[f_\kappa f_\sigma^*\right]I_{\kappa\sigma}^{\R}
	\right)~,	
\eeqn
with
\beqn
\label{eq:firstObs}
   U_\kappa^\pm 		&=& |\Amptpkappa|^2 \pm |\Amptpbarkappa|^2~, \\
   U_{\kappa\sigma}^{\pm,\R(\I)}&=& \R(\I)\left[\Amptpkappa \Amptpsigma{}^* 
				    \pm \Amptpbarkappa \Amptpbarsigma{}^*\right]~, \\
   I_\kappa			&=& \I\left[\Amptpbarkappa\Amptpkappa{}^*\right]~, \\
   I_{\kappa\sigma}^{\R}  	&=& \R\left[\Amptpbarkappa\Amptpsigma{}^* 
					    - \Amptpbarsigma\Amptpkappa{}^*\right]~, \\
\label{eq:lastObs}
   I_{\kappa\sigma}^{\I}  	&=& \I\left[\Amptpbarkappa\Amptpsigma{}^* 
					    + \Amptpbarsigma\Amptpkappa{}^*\right]~.
\eeqn

The 27 coefficients~(\ref{eq:firstObs})--(\ref{eq:lastObs}) are real-valued
parameters that multiply the $f_\kappa f_\sigma^*$ bilinears~\cite{quinnsilva}. 
They are the 
observables that are determined by the fit. Each of the coefficients 
is related in a unique way to the physically more intuitive quantities, 
like tree-level and penguin-type amplitudes, the angle $\alpha$, or
the quasi-two-body \CP and dilution parameters~\cite{rhopipaper} 
(\cf\   Section~\ref{sec:Physics}). 

The $U_\kappa^+$ coefficients are related to resonance fractions (branching
fractions and charge asymmetries), the $U_\kappa^-$ determine the relative 
abundance of the $\Bz$ decay into $\rho^+\pim$ and $\rho^-\pip$ (dilution)
and the time-dependent direct \CP asymmetries. The $I_\kappa$ measure 
mixing-induced \CP violation and are sensitive to strong phase shifts.
Finally, the $U_{\kappa\sigma}^{\pm,\R(\I)}$ and $I_{\kappa\sigma}^{\R(\I)}$
describe the interference pattern in the Dalitz plot. Their presence
distinguishes this analysis from the previous quasi-two-body one~\cite{rhopipaper}.
They represent the additional degrees of freedom that allows one to 
determine the unknown penguin pollution and the relative strong phases.
However, because 
the overlap regions of the resonances are small and because the event 
reconstruction in these regions suffers from large misreconstruction 
rates and background, a substantial data sample is needed to perform a 
fit that constrains all amplitude parameters.

The choice to fit for the $U$ and $I$ coefficients rather than  
fitting for the complex transition amplitudes and the weak phase $\alpha$ 
directly is motivated by the following technical 
simplifications: $(i)$ in contrast to the amplitudes, there is a unique 
solution for the $U$ and $I$ coefficients requiring only a single fit
to the selected data sample\footnote
{
	The parameterization~(\ref{eq:UI}) is general: the information
	on the mirror solutions (\eg, on the angle $\alpha$) that are 
	present in the transition amplitudes $\Aij$, $\Abij$ is 
	conserved.
}, 
$(ii)$ in the presence of background, the $U$ and $I$ coefficients are 
approximately Gaussian distributed, which in general is not the case 
for the amplitudes, and $(iii)$ the propagation of systematic uncertainties
and the averaging between different measurements are straightforward for 
the $U$'s and $I$'s.

We determine the quantities of interest in a subsequent least-squares
fit to the measured $U$ and $I$ coefficients.

\subsection{NORMALIZATION}

The decay rate~(\ref{eq:dt}) is used as probability density 
function (PDF) in a maximum-likelihood fit and must therefore be normalized:
\beq
 	\AmpAll \;\longrightarrow\;
	\frac{1}{\langle |\Amptp|^2 + |\Amptpbar|^2 \rangle }\AmpAll~,
\eeq
where
\beqn
\label{eq:Norm}
	\langle |\Amptp|^2 + |\Amptpbar|^2 \rangle
	\;=\;
	\sum_{\kappa\in\{+,-,0\}}  \langle|f_\kappa|^2\rangle U_\kappa^+
	\;+\;
	2\R\!\!\!\!\!\!\!\!
		\sum_{\kappa <\sigma\in\{+,-,0\}} \!\!\!\!
			\langle f_\kappa f_\sigma^*\rangle
			\left(
				U^{+,\mathrm{Re}}_{\kappa\sigma} +
				i\cdot U^{+,\mathrm{Im}}_{\kappa\sigma}
			\right)
	   ~.
\eeqn
The complex expectation values $\langle f_\kappa f_\sigma^*\rangle$ are 
obtained from high-statistics Monte Carlo integration of the Dalitz 
plot~(\ref{eq:partialWidth}), taking into account acceptance and 
resolution effects.

The normalization of the decay rate~(\ref{eq:dt}) renders the normalization
of the $U$ and $I$ coefficients arbitrary, so that we can fix one 
coefficient. By convention, we set $U_+^+\equiv1$.

For a small $\Bz\to\rho^0\pi^0$ signal, the time-dependent
\CP information from this mode is marginal. As a consequence, we 
are allowed to simplify the model by setting the coefficients of 
the sine and cosine terms that involve this mode to zero, which 
entails only a small systematic error on the result. This
reduces the number of free parameters from 26 to 16.

\subsection{THE SQUARE DALITZ PLOT}
\label{sec:SquareDP}

The signal events and also the combinatorial $\epem\to q\bar q$ ($q=u,d,s,c$) 
continuum background events populate the kinematic boundaries of the 
Dalitz plot due to the low final state masses compared to the $\Bz$ mass. 
Hence the representation 
Eq.~(\ref{eq:partialWidth}) is inadequate when one wants to use 
empirical reference shapes in a maximum-likelihood fit. Another 
practical disadvantage of the Dalitz variables is that they are 
not constant along the kinematic boundaries of the $\Btopipipi$ 
Dalitz plot, which complicates the correction of efficiency and 
resolution effects. We therefore apply the transformation
\beq
\label{eq:SqDalitzTrans}
	d\spz \,d\smz \;\longrightarrow \detJ\, d\mprime\, d\thetaprime~,
\eeq
which defines the {\em Square Dalitz plot} (SDP). The new coordinates 
are
\beq
\label{eq:SqDalitzVars}
	\mprime \;\equiv\; \frac{1}{\pi}
		\arccos\left(2\frac{\mpm - \mpmMin}{\mpmMax - \mpmMin}
			- 1
		      \right)~,
	\hspace{0.5cm}{\rm and}\hspace{0.5cm}
	\thetaprime \;\equiv\; \frac{1}{\pi}\theta_{0}~,
\eeq
where $\mpm$ is the invariant mass between the charged tracks,
$\mpmMax=\mBz - m_{\pi^0}$ and $\mpmMin=2m_{\pi^+}$ are the kinematic
limits of $\mpm$, $\theta_{0}$ is the $\rho^0$ helicity angle,
and $J$ is the Jacobian of the transformation 
that zooms into the kinematic boundaries of the Dalitz plot.
The new variables have validity ranges between 0 and 1.
The determinant of the Jacobian is given by
\beq
\label{eq:detJ}
	\detJ \;=\;	4 \,|{\bf p}^*_+||{\bf p}^*_0| \,\mpm
			\cdot 	
			\frac{\partial \mpm}{\partial \mprime}
			\cdot 	
			\frac{\partial \cos\theta_{0}}{\partial \thetaprime}~,
\eeq
where 
$|{\bf p}^*_+|=\sqrt{E^*_+ - m_{\pi^+}^2}$ and
$|{\bf p}^*_0|=\sqrt{E^*_0 - m_{\pi^0}^2}$, and where the energies 
$E^*_+$ and $E^*_0$ are in the $\pi^+\pi^-$ rest frame.

\begin{figure}[t]
  \centerline{\epsfxsize6cm\epsffile{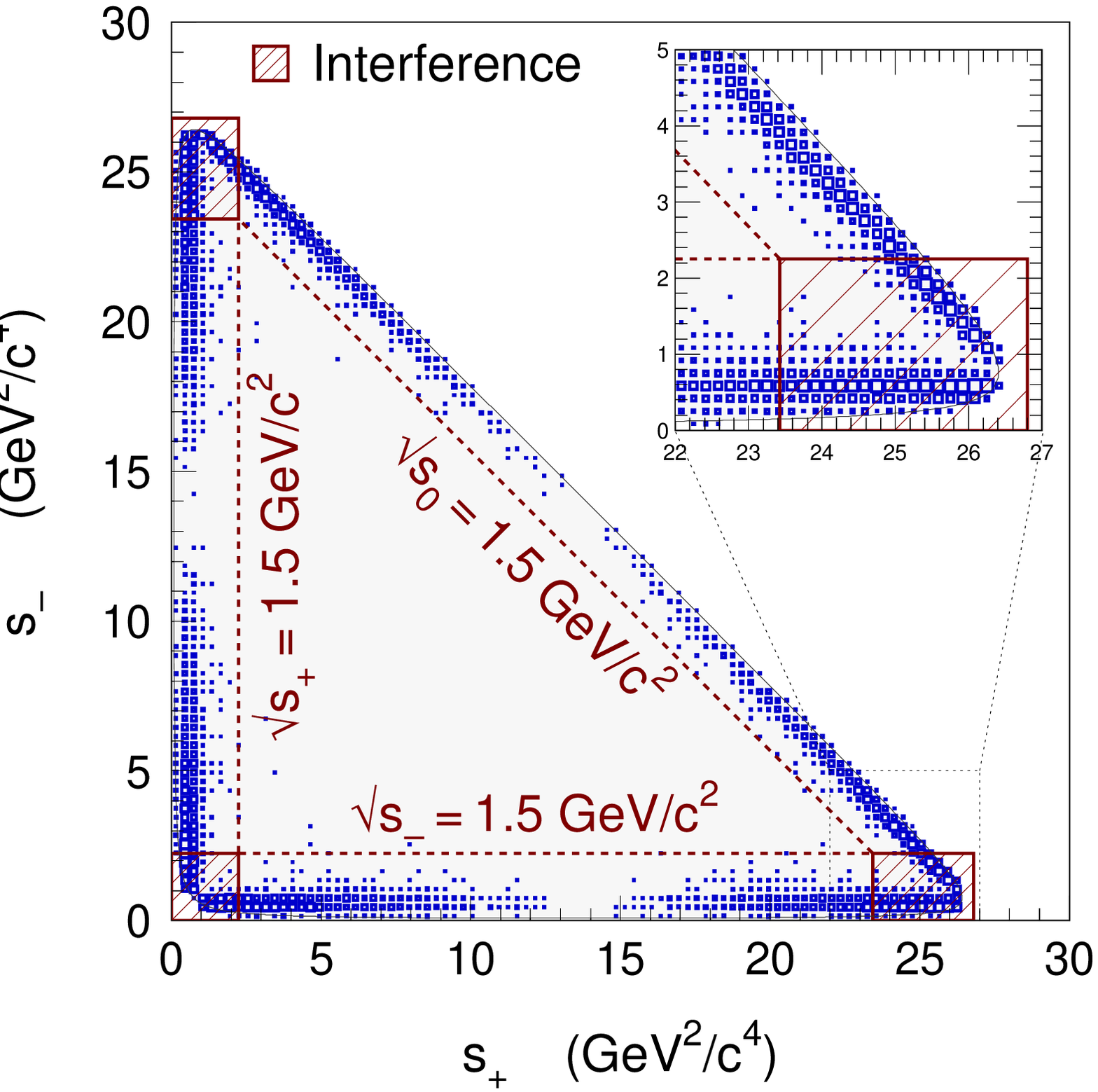}
   	      \epsfxsize4.5cm\epsffile{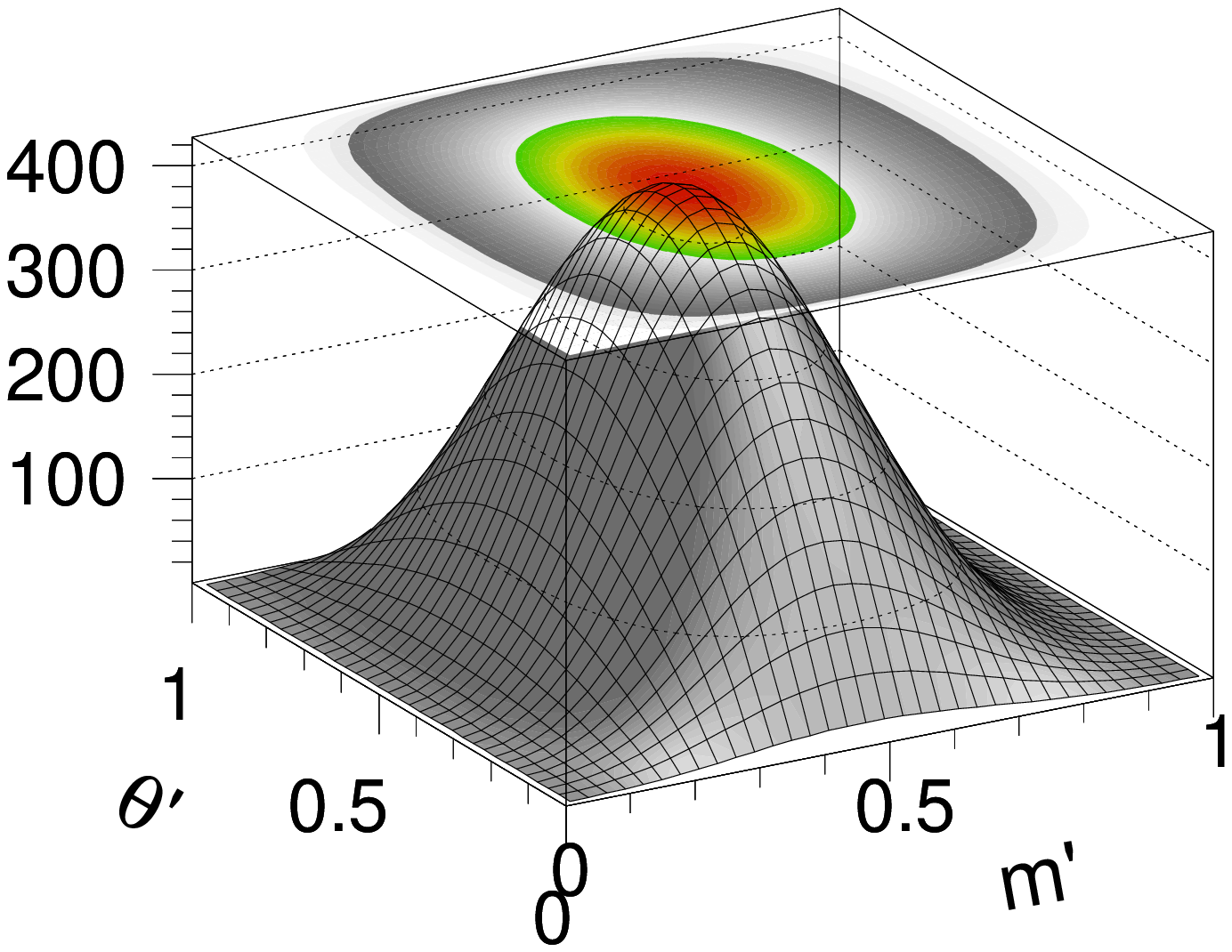}
   	      \epsfxsize6cm\epsffile{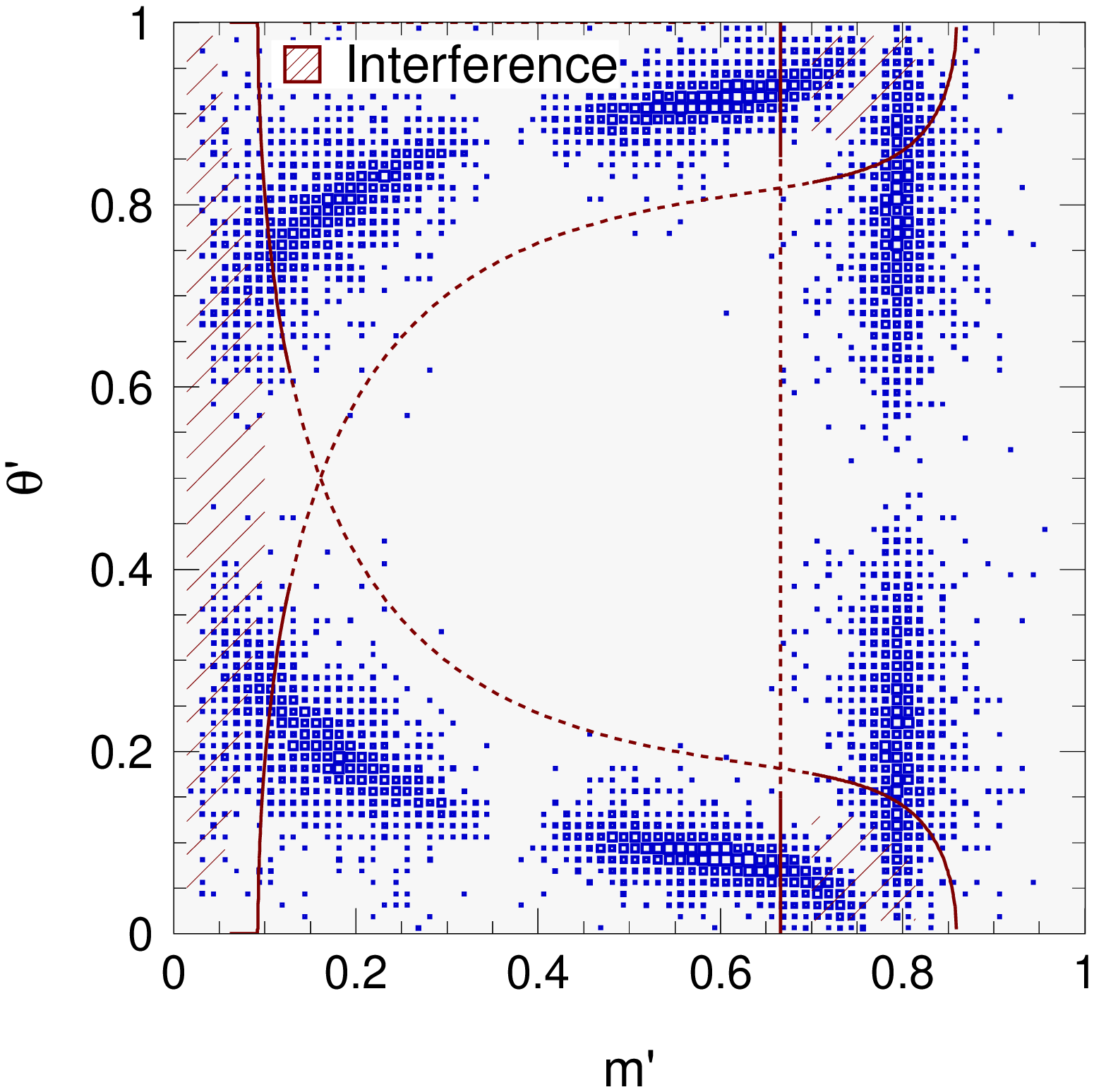}}
  \vspace{0.0cm}
  \caption[.]{\label{fig:dalitzNominal}
	Nominal (left) and square (right) 
	$\Bz\rar\pi^+\pi^-\pi^0$ Dalitz plots obtained from  
	Monte Carlo generated events without detector simulation. 
	The generating amplitudes used are $\Apm=\Amp=\Azz=1$ so that
	they interfere destructively at equal $\rho$ masses.
	The hatched areas indicate the main overlap regions between 
	the different $\rho$ bands. The contour lines in both plots
	correspond to
	$\sqrt{s_{+,-,0}}=1.5~{\rm GeV}/c^2$. The middle plot depicts
	the Jacobian determinant~(\ref{eq:detJ}) of the 
	transformation~(\ref{eq:SqDalitzTrans}). The plot shows the 
	distribution one would obtain in the square Dalitz plot for
	uniformly distributed events in the nominal Dalitz plot. }
\end{figure}

Figure~\ref{fig:dalitzNominal} shows the original (left hand plot)
and the transformed (right hand plot) Dalitz plots for 
Monte Carlo $\Btopipipi$ events generated according to 
Eqs.~(\ref{eq:amp}) and (\ref{eq:ampBar}) with equal abundance of 
all $\rho$ charges and with vanishing relative strong phase.
The plots illustrate the homogenization of the Dalitz plot
obtained after the transformation~(\ref{eq:SqDalitzTrans}). 

\section{THE \babar\ DETECTOR AND DATASET}
\label{sec:babar}

The data used in this analysis were collected with the \babar\ 
detector at the \pep2\ asymmetric-energy $e^+e^-$ storage ring at 
SLAC between October 1999 and July 2004. The sample consists of about
$192\;\mathrm{fb}^{-1}$, corresponding to $(213.1\pm2.3)\times10^{6}$ 
$B\Bbar$ pairs collected at the \FourS resonance (``on-resonance''), 
and an integrated luminosity of 11.6~\invfb collected about 40~\mev 
below the~\FourS (``off-resonance'').

A detailed description of the \babar\ detector is presented in 
Ref.~\cite{babarNim}. The tracking system used for track and vertex 
reconstruction has two main components: a silicon vertex tracker 
(SVT) and a drift chamber (DCH), both operating within a 1.5~T 
magnetic field generated by a superconducting solenoidal magnet. 
Photons are identified in an electromagnetic calorimeter (EMC) 
surrounding a detector of internally reflected Cherenkov light 
(DIRC), which associates Cherenkov photons with tracks for particle 
identification (PID). Muon candidates are identified with the
use of the instrumented flux return (IFR) of the solenoid.

\section{ANALYSIS METHOD}
\label{sec:Analysis}

The $U$ and $I$ coefficients and the $\Btopipipi$ event yield are
determined by a maximum-likelihood fit of the signal model to the 
selected candidate events. Kinematic and event shape variables 
exploiting the characteristic properties of the events are used 
in the fit to discriminate signal from background. We limit the 
size of the data sample that enters the fit by tightening the acceptance
requirements for the discriminant variables compared to similar analyses, 
because the fit model with at least 17 physical parameters is rather involved.
With the same goal and because the modeling of the distribution of 
the continuum events in the Dalitz plot is delicate, we remove the center 
of the Dalitz plot from the analysis. This requirement does not affect 
the signal.

\subsection{EVENT SELECTION AND BACKGROUND SUPPRESSION}
\label{subsec:selection}

We reconstruct $\Btopipipi$ candidates from pairs of 
oppositely-charged tracks, forming a good quality vertex, and a 
$\pi^0$ candidate. 
We use information from the tracking system, EMC, and DIRC to 
remove tracks for which the PID is consistent with the electron, kaon, 
or proton hypotheses. In addition, we require that at least one track 
has a signature in the IFR that is inconsistent with the muon 
hypothesis.
The $\pi^0$ candidate mass must satisfy $0.11<m(\gamma\gamma)<0.16\gevcc$, 
where each photon is required to have an energy greater than $50\mev$
in the laboratory frame (LAB) and to exhibit a lateral profile of energy 
deposition in the EMC consistent with an electromagnetic shower.

Two requirements are applied on the Dalitz plot. Firstly, 
the invariant mass of the two tracks $m_0$ 
must be larger than $0.52\gevcc$. This rejects
about 80\% of the $B^0 \to \KS(\to\pipi) \pi^0$ background events, which due
to the long lifetime of the $\KS$ would require a dedicated $\dt$
treatment. This cut retains 98\% (100\%) of signal $B^0 \to \rho^0\pi^0$ 
($B^0 \to \rho^\pm \pi^\mp$) events. Secondly, we remove the center 
of the Dalitz plot by requiring that at least one of the three invariant 
masses, $m_0$, $m_+$ or $m_-$, is lower than $1.5~\mathrm{GeV/c^2}$. 

A $B$-meson candidate is characterized kinematically by the energy-substituted 
mass $\mes=\lbrack{(\half s+\pvec_0\cdot\pvec_B)^2/E_0^2-\pvec_B^2}\rbrack^\half$
and energy difference $\de = E_B^*-\half\sqrt{s}$, 
where $(E_B,\pvec_B)$ and $(E_0,\pvec_0)$ are the four-vectors
of the $B$-candidate and the initial electron-positron system,
respectively. The asterisk denotes the \FourS\  frame,
and $s$ is the square of the invariant mass of the electron-positron system.  
We require $5.272 < \mes <5.288\gevcc$, which retains $81\%$
of the signal and $8\%$ of the continuum background events. 
The $\de$ resolution 
exhibits a dependence on the $\pi^0$ energy and therefore varies 
across the Dalitz plot. We account for this effect by introducing
the transformed quantity $\deprime=(2\de - \demax - \demin)/(\demax - \demin)$,
with $\deminmax(\mpm)=c_{\pm}-\left(c_{\pm}\mp\bar c\right)(\mpm/\mpmMax)^2$,
where $\mpm$ monitors the $\piz$-energy dependence. We use the values
$\bar c = 0.045\gev$, $c_{-} = -0.140\gev$, $c_{+} = 0.080\gev$,
$\mpmMax = 5.0\gev$, and require $-1<\deprime<1$. 
These settings have been obtained from Monte Carlo (MC) simulation and
are tuned to maximize the selection of correctly reconstructed over 
misreconstructed signal events. The cut retains $75\%$ ($25\%$)
of the signal (continuum).

Backgrounds arise primarily from random combinations in continuum events.
To enhance discrimination between signal and continuum, we 
use a neural network (NN) to combine four discriminating variables: 
the angles with respect to the beam axis of the $B$ momentum and $B$ thrust 
axis in the \FourS\ frame, and the zeroth and second order monomials
$L_{0,2}$ of the energy flow about the $B$ thrust axis.  The monomials
are defined by $ L_j = \sum_i p_i\times\left|\cos\theta_i\right|^j$,
where $\theta_i$ is the angle with respect to the $B$ thrust axis of
track or neutral cluster $i$, $p_i$ is its momentum, and the sum
excludes the $B$ candidate.  
The NN is trained in the signal region with off-resonance data and
simulated signal events. The final sample of signal candidates 
is selected with a cut on the NN output that retains $77\%$ ($8\%$) 
of the signal (continuum).

The time difference $\deltat$ is obtained from the measured distance between 
the $z$ positions (along the beam direction) of the $\Bz_{\tpi}$ and 
$\Bz_{\rm tag}$ decay vertices, and the boost $\beta\gamma=0.56$ of 
the \epem\ system\footnote
{
  $\deltat$ is defined as: $\deltat = \Delta z/\beta\gamma c$ 
}. 
To determine the flavor of the $\Bz_{\rm tag}$ 
we use the tagging algorithm of Ref.~\cite{BabarS2b}.
This produces four mutually exclusive tagging categories. We also 
retain untagged events in a fifth category to improve the efficiency 
of the signal selection and because these events contribute to the 
measurement of direct \CP violation. Events with multiple \B 
candidates passing the full selection occur 
in $16\%$ $(\rho^\pm\pi^\mp)$ and $9\%$ $(\rho^0\pi^0)$ 
of the cases. If the multiple candidates have different $\pi^0$'s, 
we choose the candidate with the reconstructed $\pi^0$ mass closest 
to the nominal one; if not, one candidate is selected at random. 

The signal efficiency determined from MC simulation is $24\%$ for 
$B^0 \to \rho^\pm\pi^\mp$ and $B^0 \to \rho^0\pi^0$ events, and 
$11\%$ for non-resonant $\Btopipipi$ events. 

Of the selected signal events, $22\%$ ($B^0 \to \rho^\pm\pi^\mp$), 
$13\% $ ($B^0 \to \rho^0\pi^0$), and $6\%$ (non-resonant) are 
misreconstructed, mostly due to combinatorial background from 
low-momentum tracks and photons. They concentrate in the corners
of the Dalitz plot. The fraction of misreconstructed events strongly
varies across the tagging categories.

\subsection{BACKGROUND FROM OTHER {\em B} DECAYS}

We use MC-simulated events to study the background from other $B$ 
decays. The exclusive \B-background modes are grouped into eighteen 
classes with similar kinematic and topological properties. More
than hundred decay channels have been considered of which thirty-six are 
retained in the likelihood model. The most significant ones are
$B^+ \to \rho^+\rho^0$ with longitudinal polarization ($27 \pm 18$ events 
expected), $B^+ \to \pi^+\rho^0$ 
($48 \pm 6$), $B^+ \to \pi^0\rho^+$ ($43 \pm 7$), $B^0 \to \rho^+\rho^-$ 
with longitudinal polarization ($50 \pm 10$), $B^0 \to (a_1\pi)^0$ 
($29 \pm 11$), $B^0 \to \rho^-K^+$ ($61 \pm 11$), and $B^0 \to$~higher 
kaon resonances ($6 \pm 1$). The charmed modes 
$B^0 \to D^- (\to \pi^-\pi^0) \pi^+$ and $B^0 \to \Dzb (\to \pi^+\pi^-) \pi^0$ 
contribute to the selected data sample and are considered 
in individual classes.  They do not interfere with the signal 
due to the long $D$ lifetime. We also assign classes to the modes
$B^0 \rar \Dzb(\rar K^+\pi^-)\pi^0$ and $B^0 \rar J/\psi(\rar\ell^+\ell^-)\pi^0$.
In total we expect $49\pm15$ exclusive $b\to c$ events.
Two additional classes account for inclusive neutral and charged 
$b\to c$ decays, where we expect $82 \pm 6$ and $181 \pm 9$ events, 
respectively.

%
%

\subsection{THE MAXIMUM-LIKELIHOOD FIT}
\label{subsec:ML}

We perform an unbinned extended maximum-likelihood fit to extract
the inclusive $\Btopipipi$ event yield and the $U$ and $I$ coefficients
defined in Eqs.~(\ref{eq:firstObs})--(\ref{eq:lastObs}). 
The fit uses the variables $\mes$, $\deprime$, the NN output, and the 
Dalitz plot to discriminate signal from background. The 
$\dt$ measurement allows to determine mixing-induced \CP violation
and provides additional continuum-background rejection. 

The selected on-resonance data sample is assumed to consist of signal, 
continuum-background and \B-background components, separated by the 
flavor and tagging category of the tag side \B decay. 
The signal likelihood consists of the sum of a correctly 
reconstructed (``truth-matched'', TM) component and a misreconstructed 
(``self-cross-feed'', SCF) component.

The probability density function (PDF) ${\cal P}_i^\cat$ for an
event $i$ in tagging category $\cat$ is the sum of the probability densities 
of all components, namely
\beqn
\label{eq:theLikelihood}
	{\cal P}_i^\cat
	&\equiv& 
		N_{\tpi} f^\cat_{\tpi}
		\left[ 	(1-\fscfave^\cat){\cal P}_{\tpi-\TM,i}^\cat +
			\fscfave^\cat{\cal P}_{\tpi-\SCF,i}^\cat 
		\right] 
		\nonumber\\[0.3cm]
	&&
		+\; N^\cat_{q\bar q}\frac{1}{2}
		\left(1 + \Qtagi\Atagqq\right){\cal P}_{q\bar q,i}^\cat
		\nonumber \\[0.3cm]
	&&
		+\; \sum_{j=1}^{N^{B^+}_{\rm class}}
		N_{B^+j} f^\cat_{B^+j}
		\frac{1}{2}\left(1 + \Qtagi \Atagj\right){\cal P}_{B^+,ij}^\cat
		\nonumber \\[0.3cm]
	&&
		+\; \sum_{j=1}^{N^{B^0}_{\rm class}}
		N_{B^0j} f^\cat_{B^0j}
		{\cal P}_{B^0,ij}^\cat~,
\eeqn
where: 
	$N_{\tpi}$ is the total number of $\pi^+\pi^-\pi^0$ signal events 
	the data sample;
 	$f^\cat_{\tpi}$ is the fraction of signal events that are 
       	tagged in category $\cat$;
	$\fscfave^\cat$ is the fraction of SCF events in tagging category $\cat$, 
	averaged over the Dalitz plot;
	${\cal P}_{\tpi-\TM,i}^\cat$ and ${\cal P}_{\tpi-\SCF,i}^\cat$
	are the products of PDFs of the discriminating variables used
	in tagging category $\cat$ for TM and SCF
	events, respectively; 
 	$N^\cat_{q\bar q}$ is the number of continuum events that are 
	tagged in category $\cat$;
	$\Qtagi$ is the tag flavor of the event, defined to be 
	$+1$ for a $\Bz_{\rm tag}$ and $-1$ for a $\Bzb_{\rm tag}$; 
	$\Atagqq$ parameterizes possible tag asymmetry in continuum events; 
	${\cal P}_{q\bar q,i}^\cat$ is the continuum PDF for tagging 
	category $\cat$;
	$N^{B^+}_{\rm class}$ ($N^{B^0}_{\rm class}$) is the number of 
	charged (neutral) $B$-related background classes considered in the fit;
	$N_{B^+j}$ ($N_{B^0j}$) is the number of expected events in
	the charged (neutral) $B$-background class $j$;
	$f^\cat_{B^+j}$ ($f^\cat_{B^0j}$) is the fraction of 
	charged (neutral) $B$-background events of class $j$
	that are tagged in category $\cat$;
	$\Atagj$ describes a possible tag asymmetry in the charged-$B$ background
	class $j$; 
	correlations between the tag and the position in the Dalitz plot 
	(the ``charge'') are absorbed in tag-flavor-dependent 
	Dalitz plot PDFs that are used for charged-\B and continuum
	background;
	${\cal P}_{B^+,ij}^\cat$ is the $B^+$-background PDF for tagging 
	category $\cat$ and class $j$;
	finally, ${\cal P}_{B^0,ij}^\cat$ is the neutral-$B$-background 
	PDF for tagging category $\cat$ and class $j$.

The PDFs ${\cal P}_{X}^{\cat}$ are the product of the four PDFs of the 
discriminating variables,
$x_1 = m_{ES}$, $x_2 = \deprime$, $x_3 = {\rm NN output}$, and the triplet
$x_4 = \{\mprime, \thetaprime, \deltat\}$:
\beq
\label{eq:likVars}
	{\cal P}_{X,i(j)}^{\cat} \;\equiv\; 
	\prod_{k=1}^4 P_{X,i(j)}^\cat(x_k)~.
\eeq
The extended likelihood over all tagging categories is given by
\beq
	{\cal L} \;\equiv\;  
	\prod_{\cat=1}^{5} e^{-\overline N^\cat}\,
	\prod_{i}^{N^\cat} {\cal P}_{i}^\cat~,
\eeq
where $\overline N^\cat$ is the total number of events expected in category 
$\cat$. 

A total of 39 parameters, including the inclusive signal yield and the
parameters from Eq.~(\ref{eq:dt}), are varied in the fit.

\subsubsection{\boldmath THE $\dt$ AND DALITZ PLOT PDFS}

\begin{figure}[t]
  \centerline{\epsfxsize8.2cm\epsffile{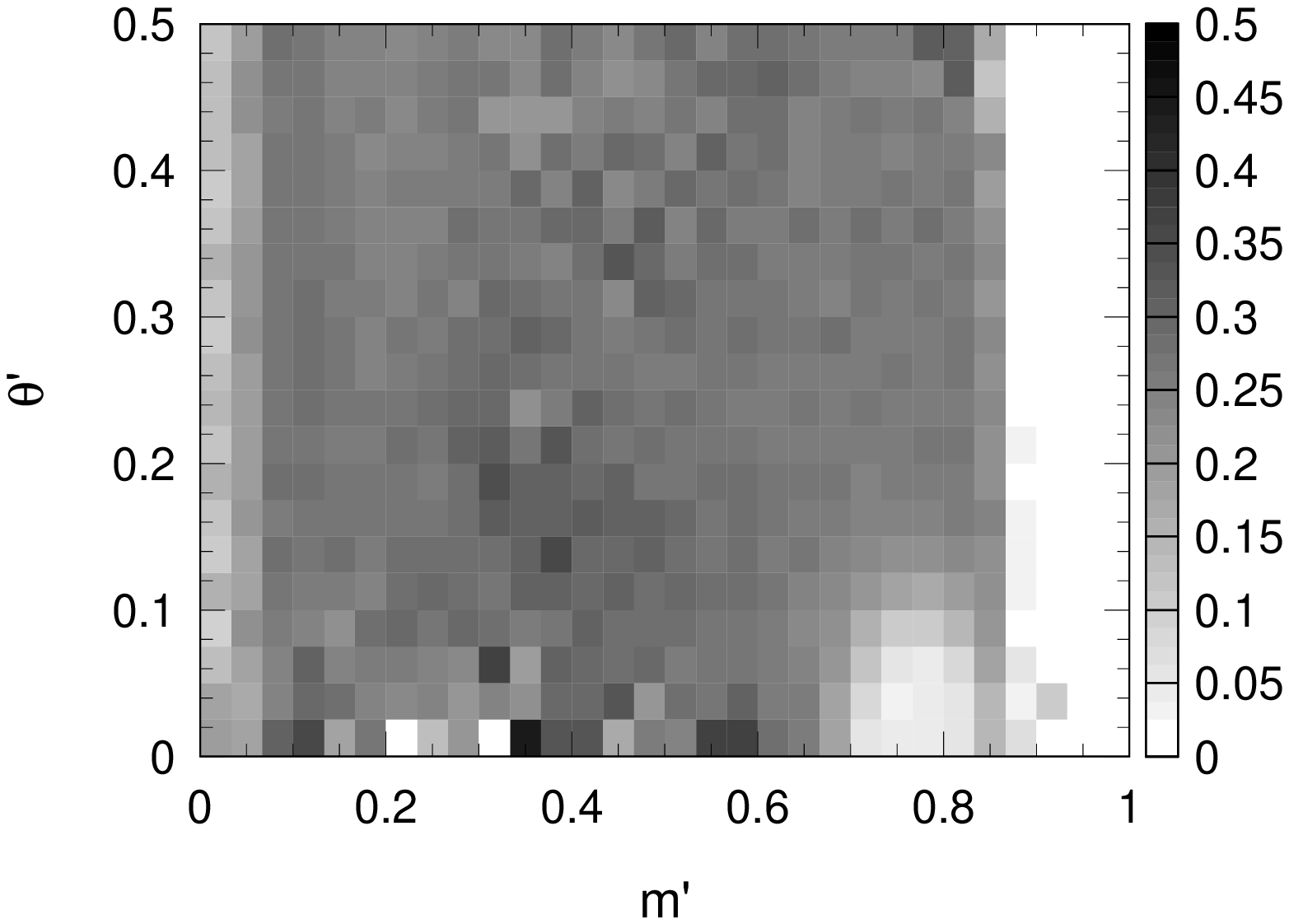}
   	      \epsfxsize8.2cm\epsffile{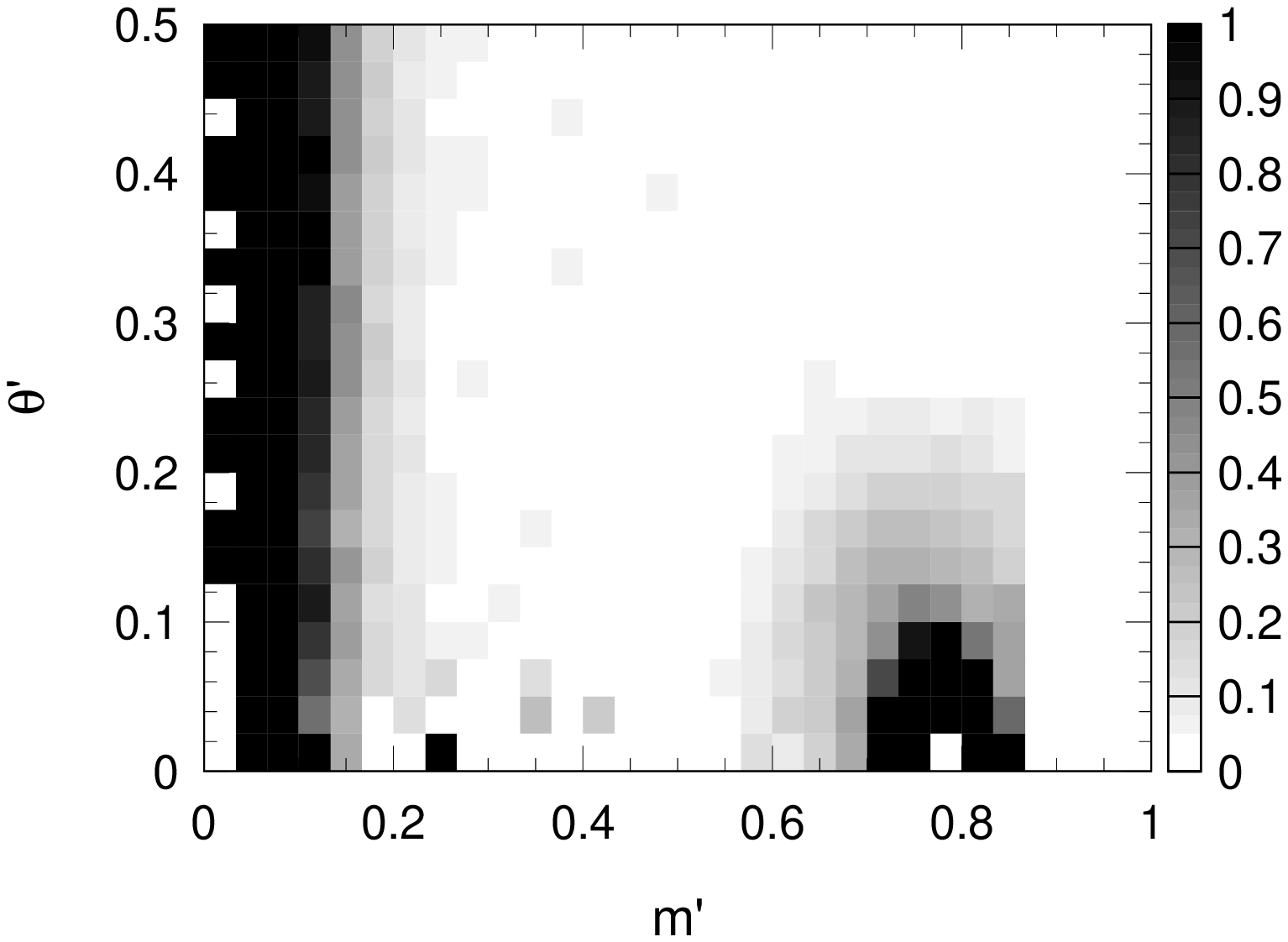}}
  \vspace{-0.4cm}
  \caption[.]{\label{fig:effscfDP}
	Selection efficiency of $\Btopipipi$ events (left) 
	and fraction of misreconstructed events (right)
	in the (symmetrized) square Dalitz plot for
	MC-simulated events.}
\end{figure}

{\bf Signal Parameterization}.
	The Dalitz plot PDFs require as input the Dalitz plot-dependent 
	relative selection efficiency, $\e=\e(\mprime,\thetaprime)$, 
	and SCF fraction, $\fscf=\fscf(\mprime,\thetaprime)$.
	Both quantities are taken from MC simulation. They are 
	given in Fig.~\ref{fig:effscfDP} (left plot for $\e$ and right 
	plot for $\fscf$), where the symmetry of the Dalitz plot has been 
	used to fold the upper $\thetaprime$ half into the lower one.
	Away from the Dalitz plot corners the efficiency is uniform, while it 
	decreases when approaching the corners, where one out of the 
	three bodies in the final state is close to rest so that the 
	acceptance requirements on the particle reconstruction become incisive.
	Combinatorial backgrounds and hence SCF fractions are large in the corners
	of the Dalitz plot due to the presence of soft neutral clusters 
	and tracks. 

	For an event~$i$, we define the time-dependent Dalitz plot PDFs
	\beqn
		P_{\tpi-\TM,i} &=&
		\varepsilon_i\,(1 - \fscfi)\,\detJi\,\AmpAll~,
		\\[0.3cm]
		P_{\tpi-\SCF,\,i} &=&
		\varepsilon_i\,\fscfi\,\detJi\,\AmpAll~,
	\eeqn	
	where $P_{\tpi-\TM,i}$ and $P_{\tpi-\SCF,\,i}$ are normalized. The 
	corresponding phase space integration involves the expectation values 	
	$\langle \varepsilon\,(1-\fscf)\,\detJ \,f^\kappa f^{\sigma*}\rangle$
	and 
	$\langle \varepsilon\,\fscf\,\detJ\, f^\kappa f^{\sigma*}\rangle$
	for TM and SCF events, where the indices $\kappa$, $\sigma$ 
	run over all resonances belonging to the signal model.
	The expectation values are model-dependent and are 
	computed with the use of MC integration over the square Dalitz plot:
	\beq
	\label{eq:normAverage}
		\langle \varepsilon\,(1-\fscf)\,\detJ\, f^\kappa f^{\sigma*}\rangle
		\;=\; \frac{\int_0^1\int_0^1 
			    \varepsilon\,(1-\fscf)\,\detJ\, f^\kappa f^{\sigma*}
			\,d\mprime d\thetaprime}
		       {\int_0^1\int_0^1 \varepsilon\,\detJ\, f^\kappa f^{\sigma*}
			\,d\mprime d\thetaprime}~,
	\eeq
	and similarly for 
	$\langle \varepsilon\,\,\detJ\, f^\kappa f^{\sigma*}\rangle$,
	where all quantities in the integrands are Dalitz plot-dependent.

	Equation~(\ref{eq:theLikelihood}) invokes the phase 
	space-averaged SCF fraction 
	$\fscfave\equiv\langle\fscf\,\detJ\, f^\kappa f^{\sigma*}\rangle$. 
	As for the PDF normalization, it is decay-dynamics-dependent.
	It has to be computed iteratively, though the remaining systematic
	uncertainty after one iteration step is small. We 
	determine the average SCF fractions separately for each tagging category 
	from MC simulation. 
	
	The width of the dominant $\rho(770)$ resonance is large compared 
	to the mass resolution for TM events (about $8\mevcc$ core Gaussian
	resolution). We can therefore neglect resolution effects in the TM 
	model.	
	Misreconstructed events	have a poor mass resolution that strongly 
	varies across the Dalitz plot. It is described in the fit by a 
	$2\times 2$-dimensional resolution function
\begin{figure}[t]
  \centerline{\epsfxsize8.2cm\epsffile{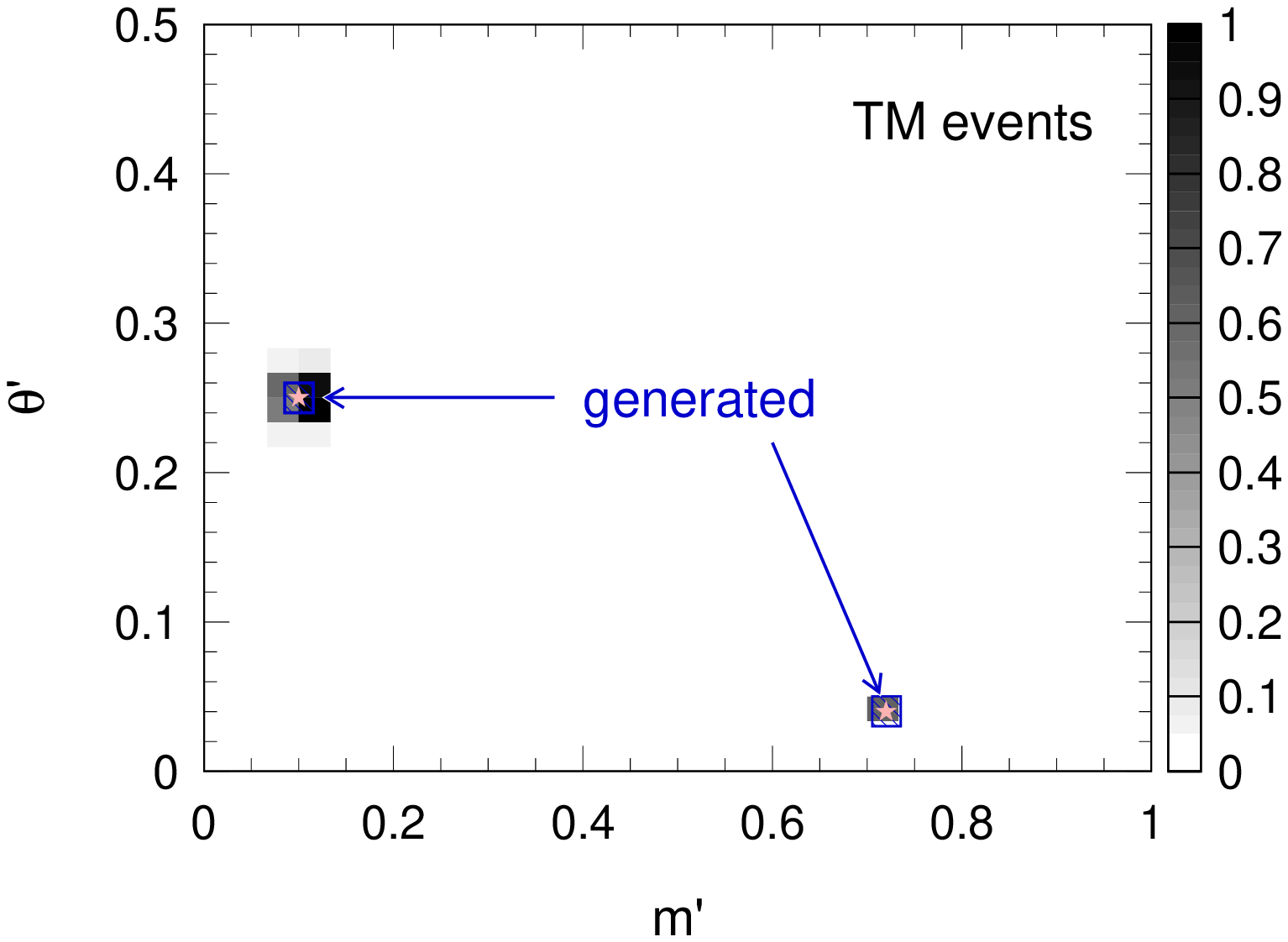}
   	      \epsfxsize8.2cm\epsffile{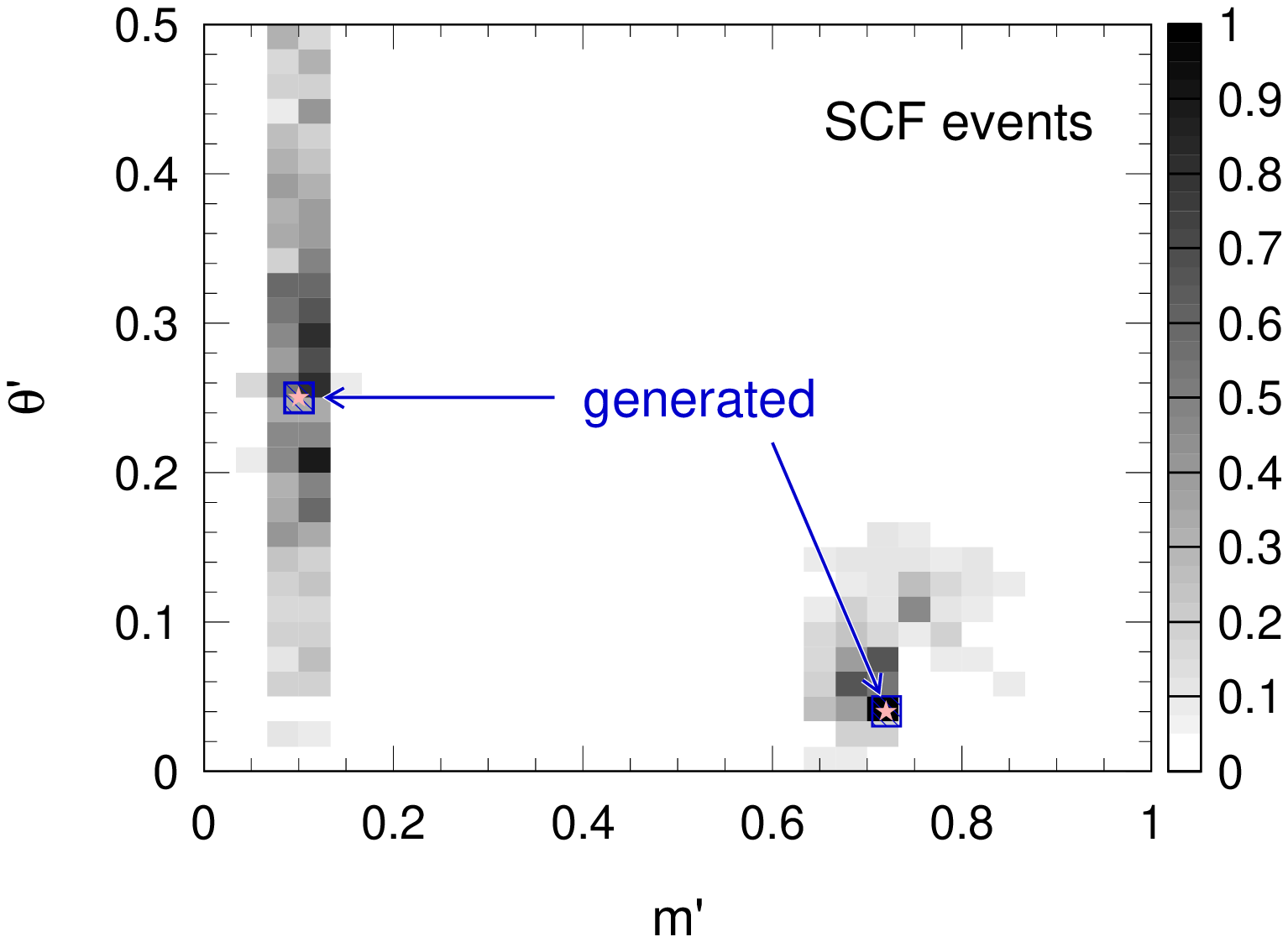}}
  \vspace{-0.4cm}
  \caption[.]{\label{fig:dpres}
	Resolution for TM (left) and SCF events (right hand plot) 
	in the square Dalitz plot for two coordinates indicated by 
	the open stars. }
\end{figure}
	\beq
	\label{eq:rscf}
		\Rscf(\mprime_r,\thetaprime_r,\mprime_t,\thetaprime_t)~,
	\eeq
	which represents the probability to reconstruct at the coordinate
	$(\mprime_r,\thetaprime_r)$ an event that has the true coordinate 
	$(\mprime_t,\thetaprime_t)$. It obeys the unitarity condition
	\beq
		\intl_0^1\intl_0^1 
		\Rscf(\mprime_r,\thetaprime_r,\mprime_t,\thetaprime_t)
		\,d\mprime_r d\thetaprime_r \;=\; 1~,
		\hspace{0.5cm} \forall\; (\mprime_t,\thetaprime_t)\in {\rm SDP}~,
	\eeq
	and is convolved with the signal model. 
	The $\Rscf$ function is obtained from MC simulation.
	Figure~\ref{fig:dpres} shows the resolution function of TM (left) 
	and SCF events (right) for two coordinates depicted by the 
	open stars.

	We use the signal model described in Section~\ref{sec:kinmeatics}. 
	It contains the dynamical information and is connected with $\dt$ via 
	the matrix element~(\ref{eq:dt}), which serves as PDF. It is diluted 
	by the effects of mistagging and the limited vertex 
	resolution~\cite{rhopipaper}. 
	The $\deltat$ resolution function for signal and \B-background 
	events is a sum of three Gaussian distributions, with parameters 
	determined by a fit to fully reconstructed $\Bz$ 
	decays~\cite{BabarS2b}. 
\\[0.3cm]\noindent
{\bf Background Parameterization}.
	The Dalitz plot- and $\dt$-dependent PDFs factorize for the 
	charged-$B$-background modes, but not (necessarily) 
	for the neutral-$B$ background due to $\BzBzb$ mixing.

 	The charged \B-background
		contribution to the likelihood~(\ref{eq:theLikelihood}) invokes 
		the parameter $\Atag$, multiplied by the tag flavor $\Qtag$ of 
		the event. In the presence of significant tag-``charge'' 
		correlation (represented by an effective 
		flavor-tag-versus-Dalitz-coordinate correlation),
		it parameterizes possible direct \CP violation in these events.
		We also use distinct square Dalitz plot PDFs for each 
		reconstructed $B$ flavor tag, and a flavor-tag-averaged PDF for 
		untagged events. The PDFs are obtained from MC simulation and are 
		described with the use of non-parametric functions.
		The $\dt$ resolution parameters are determined by a fit to fully 
		reconstructed $\Bp$ decays. For each $\Bp$-background class we adjust 
		effective lifetimes to account for the misreconstruction of the 
		event that modifies the nominal $\dt$ resolution function.

	The neutral-$B$ background is parameterized with PDFs that
		depend on the flavor tag of the event. In the case of \CP
		eigenstates, correlations between the flavor tag and the Dalitz 
		coordinate are expected to be small. However, non-\CP  eigenstates,
		such as $a_1^\pm\pi^\mp$, may exhibit such correlation. Both types 
		of decays can have direct
		and mixing-induced \CP  violation. A third type of decays
		involves charged kaons and does not exhibit mixing-induced
		\CP  violation, but usually has a strong correlation between the
		flavor tag and the Dalitz plot coordinate (the kaon charge), because 
		it consists of $B$-flavor eigenstates.
		The Dalitz plot PDFs are obtained from MC simulation and are 
		described with the use of non-parametric functions.
		For neutral $B$ background, the signal $\dt$ resolution model 
		is assumed.

	The Dalitz plot
		treatment of the continuum events is similar to the one used
		for charged-$B$ background. 
		The square Dalitz plot PDF for continuum background is 
		obtained from on-resonance events selected in the
		$\mes$ sidebands and corrected for feed-through
		from \B decays. A large number of cross checks has been 
		performed to ensure the high fidelity of the empirical shape 
		parameterization. Analytical models have been found insufficient.
		The continuum $\deltat$ distribution is parameterized as the sum of 
		three Gaussian distributions with common mean and 
		three distinct widths that scale the $\dt$ per-event error. 
		This yields six shape parameters that are determined by 
		the fit.
 		The model is motivated by the observation that 
		the $\dt$ average is independent of its error, and that the 
		$\dt$ RMS depends linearly on the $\dt$ error.

\subsubsection{PARAMETERIZATION OF THE OTHER VARIABLES}
\label{sec:likmESanddE}

	The $\mes$ distribution of TM signal events is
		parameterized by a bifurcated Crystal Ball function~\cite{PDFsCB},
		which is a combination of a one-sided Gaussian and 
		a Crystal Ball function. The mean of this function
		is determined by the fit. A non-parametric
		function is used to describe the SCF signal component.

	The $\deprime$ distribution of TM events is
		parameterized by a double Gaussian function, where
		all five parameters depend linearly on $\mpm^2$.
		Misreconstructed events are parameterized by a broad
		single Gaussian function.
		
	Both $\mes$ and $\deprime$ PDFs are parameterized by non-parametric
		functions for all $B$-background classes.

	The $\mes$ and $\deprime$ PDFs for continuum events are
		parameterized with an Argus shape function~\cite{PDFsArgus} and 
		a second order polynomial, respectively, with parameters 
		determined by the fit.

	We use non-parametric functions to empirically describe the distributions 
		of the NN outputs
		found in the MC simulation for TM and SCF signal events, 
		and for \B-background events. We distinguish tagging categories 
		for TM signal events to account for differences observed in the 
		shapes.
	
	The continuum NN distribution is parameterized by a 
		third order polynomial that is defined to be positive. 
		The coefficients of the polynomial are determined by the fit.
		Continuum events exhibit a correlation between the Dalitz plot 
		coordinate
		and the shape of the event that is exploited in the NN. The tight 
		requirement that eliminates the center of the Dalitz plot has the 
		purpose to reduce such correlation. To correct for residual effects,
		we introduce a linear dependence of the polynomial coefficients
		on the distance of the Dalitz plot coordinate to the kinematic 
		boundaries of the Dalitz plot. The parameters describing this
		dependence are determined by the fit.

\section{SYSTEMATIC STUDIES}
\label{sec:Systematics}

\begin{table}[t]
\begin{center}
\setlength{\tabcolsep}{0.0pc}
\begin{tabular*}{\textwidth}{@{\extracolsep{\fill}}lcccccccc}
\hline
&&\\[-0.3cm]
                  \rule[-6pt]{0pt}{18pt}  &       $\IM$ &       $\IP$ &      $\Uzp$ &      $\UMm$ &      $\UMp$ &      $\UPm$ &   $\UPMmIm$ &   $\UPMmRe$ \\[0.15cm]
\hline
&&\\[-0.3cm]
                 $\dmd$ and $\tau_{\Bz}$  &  $ 0.003$   &  $ 0.001$   &  $ 0.000$   &  $ 0.004$   &  $ 0.000$   &  $ 0.001$   &  $ 0.003$   &  $ 0.007$   \\
                      Signal description  &  $ 0.005$   &  $ 0.004$   &  $ 0.001$   &  $ 0.011$   &  $ 0.013$   &  $ 0.010$   &  $ 0.042$   &  $ 0.056$   \\
                                 Tagging  &  $ 0.003$   &  $ 0.002$   &  $ 0.001$   &  $ 0.010$   &  $ 0.001$   &  $ 0.010$   &  $ 0.042$   &  $ 0.019$   \\
                            Signal model  &  $ 0.004$   &  $ 0.005$   &  $ 0.011$   &  $ 0.013$   &  $ 0.004$   &  $ 0.020$   &  $ 0.202$   &  $ 0.154$   \\
                          $B$ Background  &  $ 0.011$   &  $ 0.013$   &  $ 0.007$   &  $ 0.029$   &  $ 0.015$   &  $ 0.035$   &  $ 0.118$   &  $ 0.071$   \\
               Continuum parametrization  &  $ 0.004$   &  $ 0.004$   &  $ 0.040$   &  $ 0.002$   &  $ 0.012$   &  $ 0.004$   &  $ 0.027$   &  $ 0.031$   \\
      Fixing 10 $\rho^0\piz$ parameters   &  $ 0.003$   &  $ 0.003$   &  $ 0.025$   &  $ 0.019$   &  $ 0.007$   &  $ 0.039$   &  $ 0.042$   &  $ 0.022$   \\
                                Fit Bias  &  $ 0.015$   &  $ 0.014$   &  $ 0.006$   &  $ 0.022$   &  $ 0.015$   &  $ 0.020$   &  $ 0.177$   &  $ 0.161$   \\[0.1cm]
\hline
              \rule[-6pt]{0pt}{18pt} Sum  &  $ 0.021$   &  $ 0.021$   &  $ 0.050$   &  $ 0.046$   &  $ 0.029$   &  $ 0.061$   &  $ 0.303$   &  $ 0.244$   \\[0.1cm]
\hline
\end{tabular*}
\vspace{1\baselineskip}

\setlength{\tabcolsep}{0.0pc}
\begin{tabular*}{\textwidth}{@{\extracolsep{\fill}}lcccccccc}
\hline
&&\\[-0.3cm]
                  \rule[-6pt]{0pt}{18pt}  &   $\UPMpIm$ &   $\UPMpRe$ &    $\IPMIm$ &    $\IPMRe$ &   $\UPzpIm$ &   $\UPzpRe$ &   $\UMzpIm$ &   $\UMzpRe$ \\[0.15cm]
\hline
&&\\[-0.3cm]
                 $\dmd$ and $\tau_{\Bz}$  &  $ 0.002$   &  $ 0.001$   &  $ 0.010$   &  $ 0.041$   &  $ 0.001$   &  $ 0.001$   &  $ 0.001$   &  $ 0.001$   \\
                      Signal description  &  $ 0.014$   &  $ 0.043$   &  $ 0.108$   &  $ 0.063$   &  $ 0.013$   &  $ 0.029$   &  $ 0.048$   &  $ 0.016$   \\
                                 Tagging  &  $ 0.009$   &  $ 0.011$   &  $ 0.062$   &  $ 0.041$   &  $ 0.004$   &  $ 0.006$   &  $ 0.005$   &  $ 0.007$   \\
                            Signal model  &  $ 0.079$   &  $ 0.137$   &  $ 0.629$   &  $ 0.333$   &  $ 0.129$   &  $ 0.096$   &  $ 0.107$   &  $ 0.161$   \\
                          $B$ Background  &  $ 0.065$   &  $ 0.038$   &  $ 0.159$   &  $ 0.179$   &  $ 0.012$   &  $ 0.040$   &  $ 0.038$   &  $ 0.038$   \\
               Continuum parametrization  &  $ 0.031$   &  $ 0.034$   &  $ 0.097$   &  $ 0.099$   &  $ 0.019$   &  $ 0.059$   &  $ 0.012$   &  $ 0.028$   \\
      Fixing 10 $\rho^0\piz$ parameters   &  $ 0.019$   &  $ 0.010$   &  $ 0.024$   &  $ 0.004$   &  $ 0.110$   &  $ 0.069$   &  $ 0.102$   &  $ 0.000$   \\
                                Fit Bias  &  $ 0.078$   &  $ 0.076$   &  $ 0.221$   &  $ 0.240$   &  $ 0.051$   &  $ 0.048$   &  $ 0.055$   &  $ 0.047$   \\[0.1cm]
\hline
              \rule[-6pt]{0pt}{18pt} Sum  &  $ 0.135$   &  $ 0.171$   &  $ 0.704$   &  $ 0.467$   &  $ 0.179$   &  $ 0.149$   &  $ 0.170$   &  $ 0.175$   \\[0.1cm]
\hline
\end{tabular*}

\vspace{-0.2cm}
\caption{ \label{tab:systematics}
        Summary of systematic uncertainties.}
\end{center}
\end{table}
The contributions to the systematic error on the signal parameters are 
summarized in Table~\ref{tab:systematics}. 

	The uncertainties associated with $\dmd$ and $\tau$ are 
	estimated by varying these parameters within the uncertainties 
	on the world average~\cite{PDG}. 

	The systematic effects due to the signal PDFs (``Signal description'' 
	field in Table~\ref{tab:systematics}) comprise uncertainties in the 
	PDF parameterization, the treatment of misreconstructed events, the 
	tagging performance, and the modeling of the signal contributions.

	When the signal PDFs are determined from fits to a control sample 
	of fully reconstructed \B decays to exclusive final states with 
	charm, the uncertainties are obtained by varying the parameters 
	within the statistical uncertainties. 
	In other cases, the dominant parameters have been 
	left free to vary in the fit, and the differences observed in these
	fits are taken as systematic errors. 

	The average fraction of misreconstructed signal events predicted by the MC 
	simulation has been verified with fully reconstructed $\B\to D\rho$
	events~\cite{rhopipaper}. No significant differences between data and 
	the simulation were found. We vary $\fscfave$ for all tagging categories 
	by relative $25\%$ to estimate the systematic uncertainty (contained
	in ``Signal description'' in Table~\ref{tab:systematics}).

	Tagging efficiencies, dilutions and biases for signal events
	are varied within their	experimental uncertainties.

	The most important contribution to the systematic uncertainty stems
	from the signal modeling of the Dalitz plot dynamics. We vary the mass
	and width of the $\rho(770)$ and $\rho(1450)$ within ranges that 
	exceed twice the errors found for these parameters in the fits to 
	$\tau$ and $\epem$ data~\cite{taueeref}. Since some of the $U$ and 
	$I$ coefficients exhibit significant dependence on the $\rho(1450)$ 
	contribution, we leave its amplitude (phase and 
	fraction) free to vary in the nominal fit. We vary the relative amount of 
	$\rho(1700)$ by $30\%$ with respect to the nominal model, and its phase
	by $8^\circ$, to assign a systematic error. We have performed a fit 
	where the $\rho(1700)$ amplitude parameters (magnitude and phase) are
	free to vary. We find results that 
	are in agreement with the nominal model. The variations for the 
	$U$ and $I$ coefficients observed in this fit compared to the nominal 
	one are smaller than the systematic uncertainties we assign due to the 
	$\rho(1700)$ amplitude uncertainty.

	To estimate the contribution from non-resonance $\Btopipipi$ events, we have 
	performed an independent analysis where we apply the contrary of
	the Dalitz plot requirement that is used in the nominal analysis;
	to remove the $\rho$ signal we retain only those events for which 
	the minimum invariant mass exceeds $1.5\gevcc$. For simplicity, 
	we assume a uniform Dalitz distribution
	for the non-resonance events. The fit finds no non-resonant
	events so that we can determine a preliminary upper 
	limit of $1.4\times10^{-6}$ at $90\%$ confidence level (statistical 
	errors only). According to this limit, we add simulated non-resonant 
	events to the nominal data sample to estimate the systematic uncertainty.
	We have also searched for the presence of $\Bz\to f_0(980)\piz$ events
	without finding evidence for a signal.

	A major source of systematic uncertainty is the $B$-background model. 
	The expected event yields from the background modes are varied according 
	to the uncertainties in the measured or estimated branching fractions.  
	Since $B$-background modes may exhibit  \CP violation, the corresponding 
	parameters are varied within appropriate uncertainty ranges. As is done
	for the signal PDFs, we vary the $\dt$ resolution parameters and
	the flavor-tagging parameters within their uncertainties and assign
	the differences observed in the on-resonance data fit with respect to
	nominal fit as systematic errors.

	The parameters for 
	the continuum events are determined by the fit. No additional systematic 
	uncertainties are assigned to them. An exception to this is the Dalitz 
	plot PDF: to estimate the systematic
	uncertainty from the \mes sideband extrapolation, we select large 
	samples of off-resonance data by loosening the requirements on \de and 
	the NN. We compare the distributions of $\mprime$ and $\thetaprime$ 
	between the \mes sideband and the signal region. No significant 
	differences are found. We assign as systematic error the effect seen when
	weighting the continuum Dalitz plot PDF by the ratio of both data 
	sets. This effect is mostly statistical in origin. To account for
	possible inaccuracies in the empirical parameterization, we add 
	test components with floating event yields to the fit that consist of 
	continuum-background-like reference distributions in all fit variables
	but the Dalitz plot, for which signal-like distributions are used.
	In addition, we leave the \B-background classes free to vary in the fit.
	This allows the fit to absorb events that, due to possible problems 
	in the continuum Dalitz plot description, would bias the signal yield.
	The shifts in the signal parameters observed when adding these test
	components are taken as systematic uncertainties (``continuum
	parameterization'' in Table~\ref{tab:systematics}). It leads to
	significant effects in the $\rho^0\piz$ region of the Dalitz plot.

	We assess the dependence of the results on whether all the 
	27 $U$ and $I$ coefficients are free in the fit or only the 16 most 
	significant ones. To study this effect, we generate MC samples with 
	values for the $U$ 
	and $I$ coefficients according to our nominal 16 parameter fit result.
	We then perform a full amplitude fit to these 16 coefficients 
	(\cf\  Section~\ref{sec:Physics}) and compute from the best fit the 
	expected values for the missing coefficients. Monte Carlo samples
	are generated with the use of all 27 coefficients, which are fit
	with the nominal model where 10 coefficients are set to zero. The 
	observed systematic effect on the measured $U$ and $I$ coefficients
	depend on the branching fraction for $\Bz\to\rho^0\piz$ for which 
	we use our upper bound~\cite{rho0pi0paper} (which is in agreement with 
	the finding in this analysis).

	Finally, 
	to validate the fitting tool, we perform fits on large MC samples with 
	the measured proportions of signal, continuum and $B$-background events.
	No significant biases are observed in these fits. The statistical
	uncertainties on the fit parameters are taken as systematic uncertainties
	(``Fit bias'').

The systematic errors for the parameters that measure interference
effects are dominated by the uncertainty in the signal model, mainly
the tail description of the $\rho$ resonance. For the other parameters,
the uncertainty on the fit bias and the \B-background contamination 
are important.

\section{FIT RESULTS}
\label{sec:fitResults}

\begin{table}[t]
\begin{center}
\setlength{\tabcolsep}{0.3pc}
\begin{tabular*}{\textwidth}{@{\extracolsep{\fill}}llc}
\hline
&&\\[-0.3cm]
Parameter   & Description & Result \\[0.15cm]
\hline
&&\\[-0.3cm]
\rule[-1.7mm]{0mm}{5mm}$I_-$
     & Coefficient of $|f_-|^2\sin(\dmd\dmt)$ & $-0.19\pm0.11\pm0.02$ \\
\rule[-1.7mm]{0mm}{5mm}$I_+$           
     & Coefficient of $|f_+|^2\sin(\dmd\dmt)$ & $\phantom{-}0.06\pm0.11\pm0.02$ \\
\rule[-1.7mm]{0mm}{5mm}$U_0^+$
     & Coefficient of $|f_0|^2$               & $\phantom{-}0.16\pm0.05\pm0.05$ \\
\rule[-1.7mm]{0mm}{5mm}$U_-^-$         
     & Coefficient of $|f_-|^2\cos(\dmd\dmt)$ & $\phantom{-}0.22\pm0.16\pm0.05$  \\
\rule[-1.7mm]{0mm}{5mm}$U_-^+$         
     & Coefficient of $|f_-|^2$               & $\phantom{-}1.19\pm0.12\pm0.03$  \\
\rule[-1.7mm]{0mm}{5mm}$U_+^-$         
     & Coefficient of $|f_+|^2\cos(\dmd\dmt)$ & $\phantom{-}0.50\pm0.17\pm0.06$  \\
\rule[-1.7mm]{0mm}{5mm}$U_{+-}^{-,\I}$ 
     & Coefficient of $\I[f_+f_-^*]\cos(\dmd\dmt)$ & $\phantom{-}0.3\ph{0}\pm1.4\ph{0}\pm0.3\ph{0}$  \\
\rule[-1.7mm]{0mm}{5mm}$U_{+-}^{-,\R}$ 
     & Coefficient of $\R[f_+f_-^*]\cos(\dmd\dmt)$ & $\phantom{-}2.0\ph{0}\pm1.2\ph{0}\pm0.2\ph{0}$   \\
\rule[-1.7mm]{0mm}{5mm}$U_{+-}^{+,\I}$ 
     & Coefficient of $\I[f_+f_-^*]$ & $\phantom{-}0.16\pm0.70\pm0.14$   \\
\rule[-1.7mm]{0mm}{5mm}$U_{+-}^{+,\R}$ 
     & Coefficient of $\R[f_+f_-^*]$ & $-0.26\pm0.65\pm0.17$    \\
\rule[-1.7mm]{0mm}{5mm}$I_{+-}^{\I}$   
     & Coefficient of $\I[f_+f_-^*]\sin(\dmd\dmt)$ & $-5.2\ph{0}\pm1.9\ph{0}\pm0.7\ph{0}$  \\
\rule[-1.7mm]{0mm}{5mm}$I_{+-}^{\R}$   
     & Coefficient of $\R[f_+f_-^*]\sin(\dmd\dmt)$ & $-0.3\ph{0}\pm2.0\ph{0}\pm0.5\ph{0}$ \\
\rule[-1.7mm]{0mm}{5mm}$U_{+0}^{+,\I}$ 
     & Coefficient of $\I[f_+f_0^*]$ & $\phantom{-}0.25\pm0.35\pm0.18$ \\
\rule[-1.7mm]{0mm}{5mm}$U_{+0}^{+,\R}$ 
     & Coefficient of $\R[f_+f_0^*]$ & $-0.34\pm0.39\pm0.15$ \\
\rule[-1.7mm]{0mm}{5mm}$U_{-0}^{+,\I}$ 
     & Coefficient of $\I[f_-f_0^*]$ & $\phantom{-}0.34\pm0.43\pm0.17$ \\
\rule[-1.7mm]{0mm}{5mm}$U_{-0}^{+,\R}$ 
     & Coefficient of $\R[f_-f_0^*]$ & $-0.98\pm0.44\pm0.18$ \\[0.15cm]
\hline
\end{tabular*}
\caption{Fit results for the $U$ and $I$ coefficients. 
	The errors given are statistical (first) and systematic (second).
	The free normalization parameter $U_+^+$ is fixed to 1.}
\label{tab:results}
\end{center}
\end{table}
\begin{table}[t]
\begin{center}
\setlength{\tabcolsep}{0.0pc}
{\small
\begin{tabular*}{\textwidth}{@{\extracolsep{\fill}}lrrrrrrrrrrrrrrr}
\hline
&&\\[-0.3cm]
             \rule[-6pt]{0pt}{18pt}          &     $\IM$   &     $\IP$   &    $\Uzp$   &    $\UMm$   &    $\UMp$   &    $\UPm$   & $\UPMmIm$   & $\UPMmRe$  &   $\UPMpIm$   & $\UPMpRe$   &  $\IPMIm$   &  $\IPMRe$   & $\UPzpIm$   & $\UPzpRe$   & $\UMzpIm$   \\[0.15cm]
\hline                                                                                                 &&\\[-0.3cm]
 $\IM      $&$ 1.00 $&$ $&$ $&$ $&$ $&$ $&$ $&$ $&$ $&$ $&$ $&$ $&$ $&$ $&$ $\\
 $\IP      $&$ 0.00 $&$ 1.00 $&$ $&$ $&$ $&$ $&$ $&$ $&$ $&$ $&$ $&$ $&$ $&$ $&$ $\\
 $\Uzp     $&$ 0.16 $&$ 0.21 $&$ 1.00 $&$ $&$ $&$ $&$ $&$ $&$ $&$ $&$ $&$ $&$ $&$ $&$ $\\
 $\UMm     $&$ 0.03 $&$ 0.01 $&$ 0.01 $&$ 1.00 $&$ $&$ $&$ $&$ $&$ $&$ $&$ $&$ $&$ $&$ $&$ $\\
 $\UMp     $&$ 0.01 $&$ 0.01 $&$-0.05 $&$ 0.00 $&$ 1.00 $&$ $&$ $&$ $&$ $&$ $&$ $&$ $&$ $&$ $&$ $\\
 $\UPm     $&$-0.06 $&$-0.03 $&$ 0.01 $&$-0.03 $&$-0.06 $&$ 1.00 $&$ $&$ $&$ $&$ $&$ $&$ $&$ $&$ $&$ $\\
 $\UPMmIm  $&$ 0.07 $&$-0.04 $&$ 0.07 $&$-0.10 $&$-0.05 $&$-0.04 $&$ 1.00 $&$ $&$ $&$ $&$ $&$ $&$ $&$ $&$ $\\
 $\UPMmRe  $&$-0.02 $&$-0.01 $&$-0.04 $&$ 0.02 $&$ 0.01 $&$ 0.01 $&$-0.05 $&$ 1.00 $&$ $&$ $&$ $&$ $&$ $&$ $&$ $\\
 $\UPMpIm  $&$-0.10 $&$-0.01 $&$-0.09 $&$ 0.05 $&$ 0.13 $&$ 0.13 $&$-0.05 $&$ 0.15 $&$ 1.00 $&$ $&$ $&$ $&$ $&$ $&$ $\\
 $\UPMpRe  $&$-0.03 $&$ 0.02 $&$-0.02 $&$ 0.02 $&$-0.08 $&$ 0.07 $&$-0.01 $&$-0.01 $&$-0.03 $&$ 1.00 $&$ $&$ $&$ $&$ $&$ $\\
 $\IPMIm   $&$ 0.09 $&$-0.05 $&$ 0.06 $&$-0.10 $&$ 0.04 $&$-0.00 $&$ 0.27 $&$-0.04 $&$-0.14 $&$-0.10 $&$ 1.00 $&$ $&$ $&$ $&$ $\\
 $\IPMRe   $&$-0.02 $&$ 0.00 $&$ 0.01 $&$ 0.06 $&$ 0.03 $&$-0.02 $&$-0.02 $&$-0.04 $&$ 0.05 $&$-0.03 $&$-0.01 $&$ 1.00 $&$ $&$ $&$ $\\
 $\UPzpIm  $&$-0.07 $&$-0.00 $&$-0.18 $&$ 0.03 $&$ 0.03 $&$-0.01 $&$-0.11 $&$ 0.08 $&$ 0.06 $&$ 0.02 $&$-0.10 $&$ 0.12 $&$ 1.00 $&$ $&$ $\\
 $\UPzpRe  $&$-0.00 $&$-0.01 $&$ 0.04 $&$-0.19 $&$ 0.01 $&$ 0.00 $&$ 0.13 $&$-0.00 $&$-0.06 $&$-0.04 $&$ 0.13 $&$ 0.01 $&$-0.16 $&$ 1.00 $&$ $\\
 $\UMzpIm  $&$ 0.02 $&$-0.01 $&$ 0.13 $&$ 0.01 $&$-0.05 $&$ 0.01 $&$ 0.03 $&$-0.01 $&$-0.04 $&$ 0.04 $&$ 0.00 $&$-0.15 $&$-0.19 $&$-0.00 $&$ 1.00 $\\
 $\UMzpRe  $&$-0.05 $&$-0.05 $&$-0.05 $&$ 0.01 $&$ 0.01 $&$ 0.04 $&$-0.06 $&$ 0.03 $&$ 0.13 $&$ 0.01 $&$-0.07 $&$ 0.12 $&$ 0.14 $&$-0.05 $&$-0.01 $
\\[0.15cm]\hline			      
\end{tabular*} 
}

\end{center}
\vspace{-0.6cm}
\caption{\label{tab:corrmat}
	Correlation matrix for the $U$ and $I$ coefficients 
	obtained including statistical and systematic errors.}
\end{table}
\begin{figure}[p]
  \centerline{  \epsfxsize8.2cm\epsffile{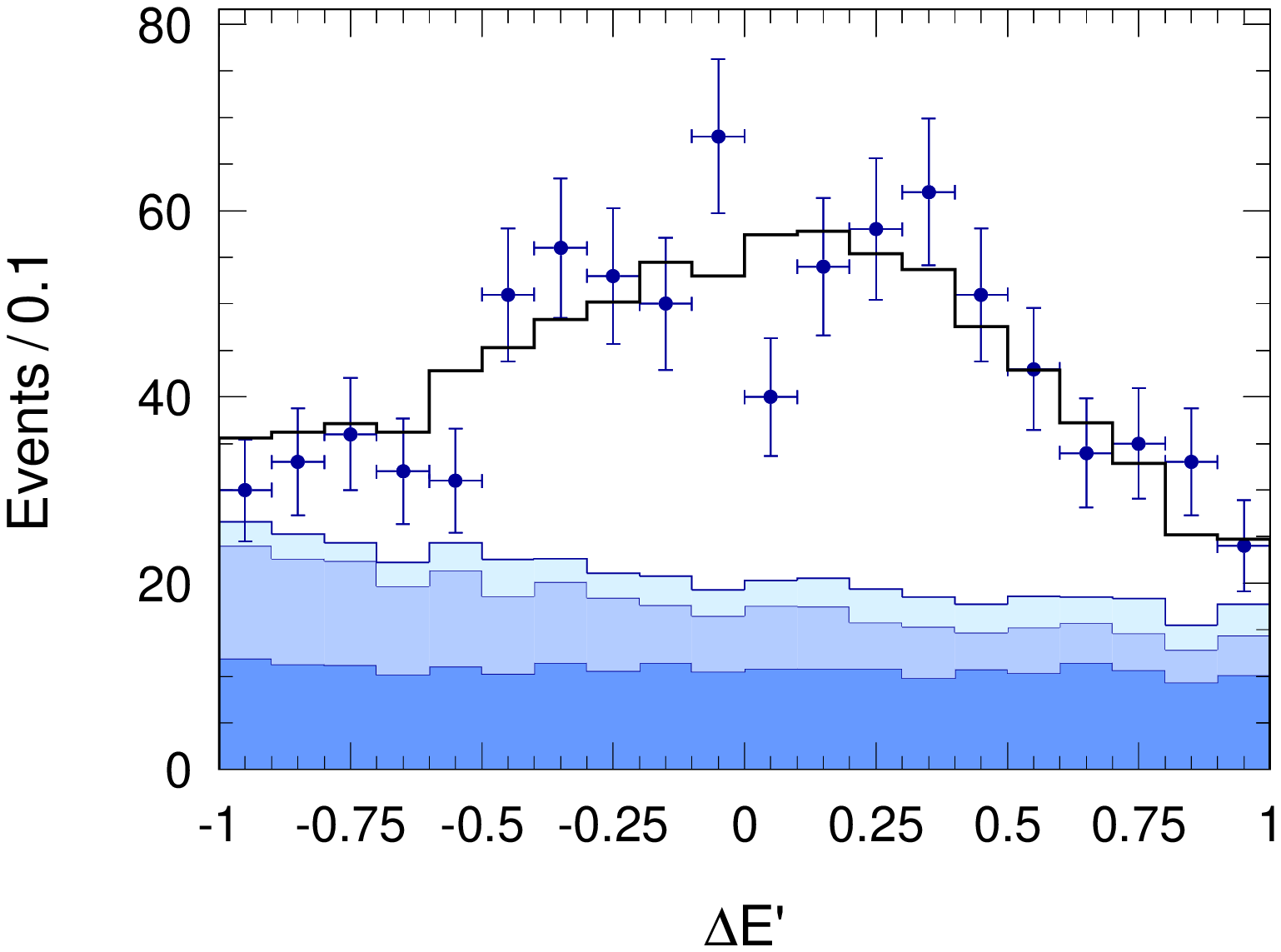}
                \epsfxsize8.2cm\epsffile{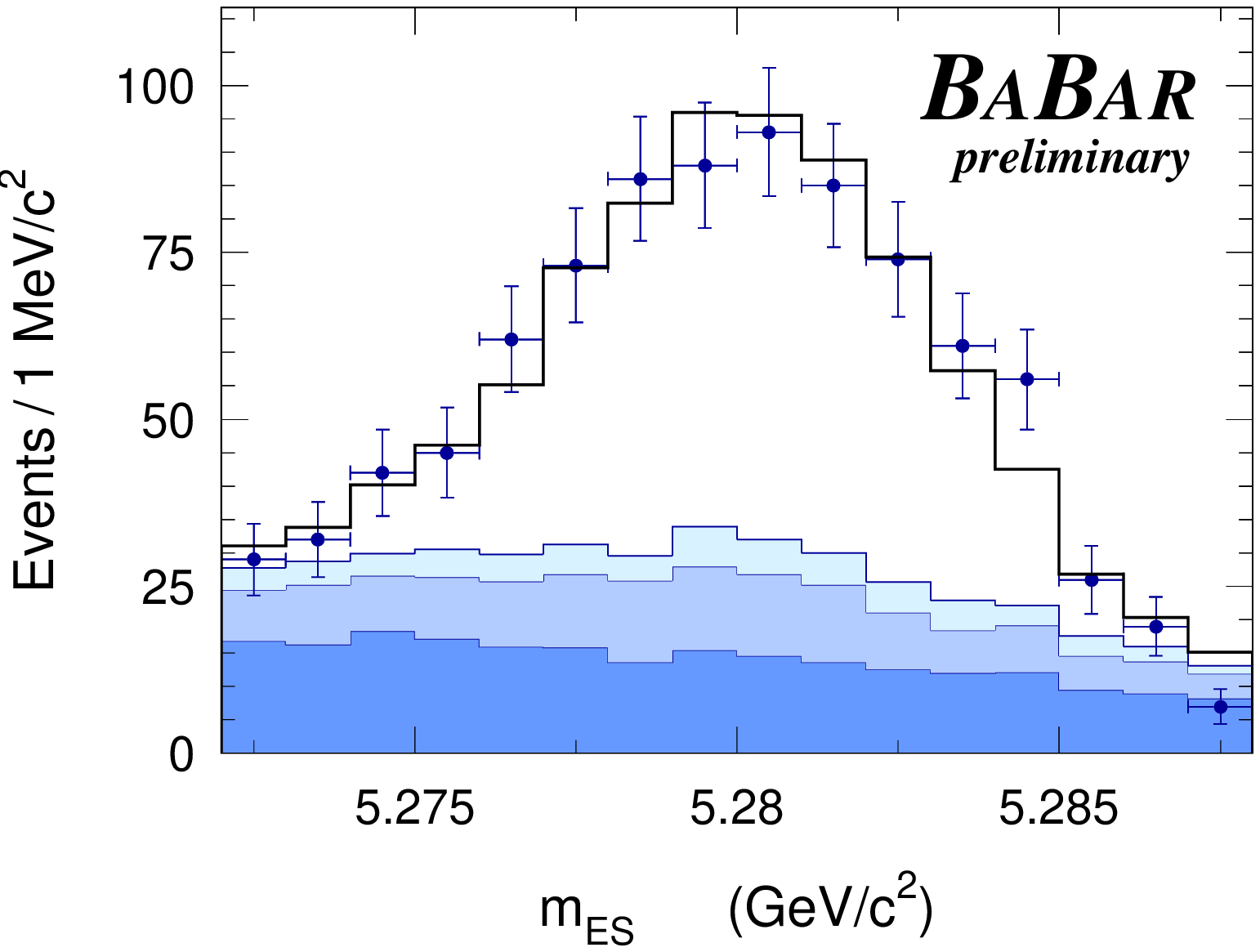}}
  \vspace{-0.1cm}
  \centerline{  \epsfxsize8.2cm\epsffile{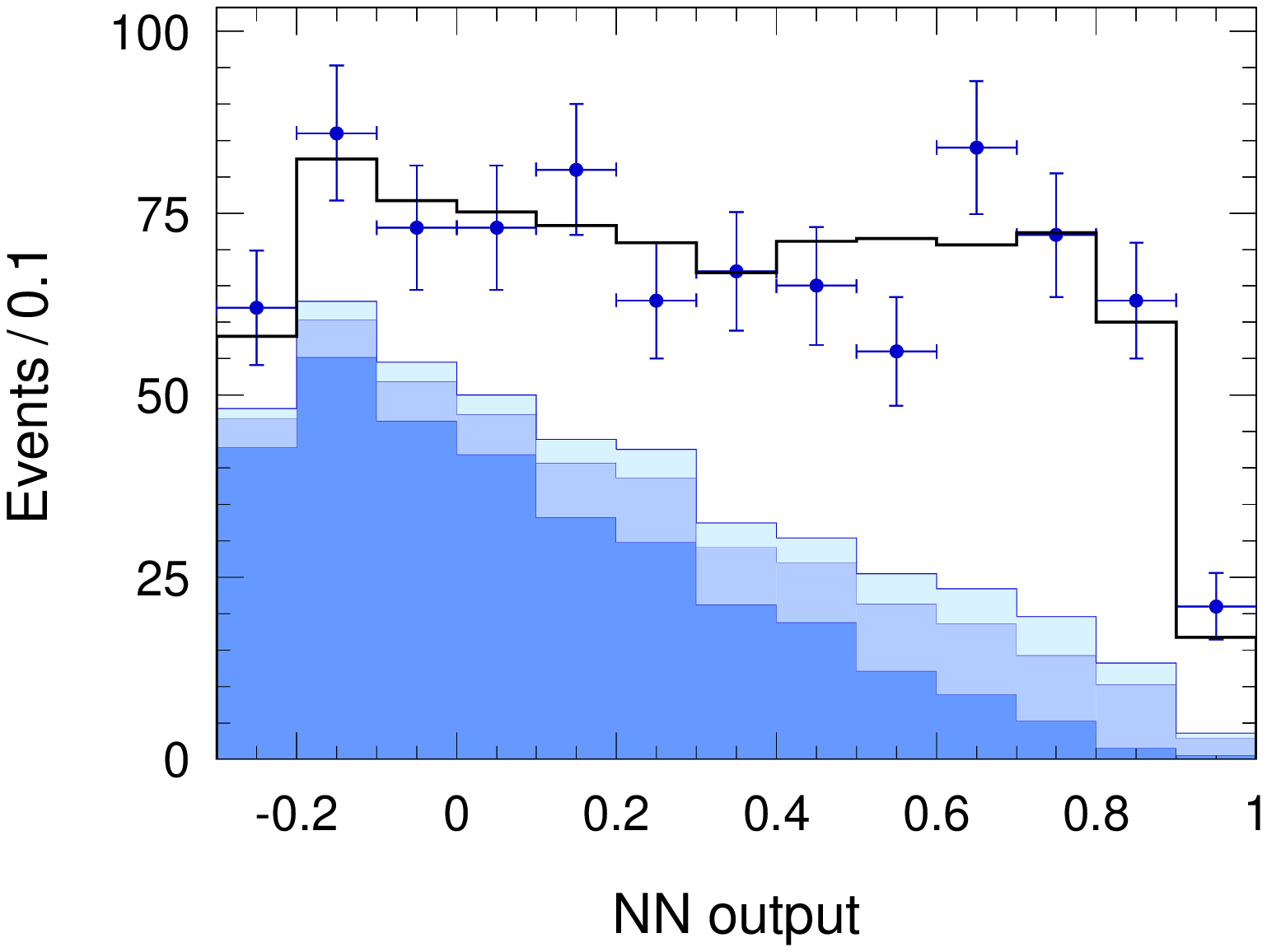} 
                \epsfxsize8.2cm\epsffile{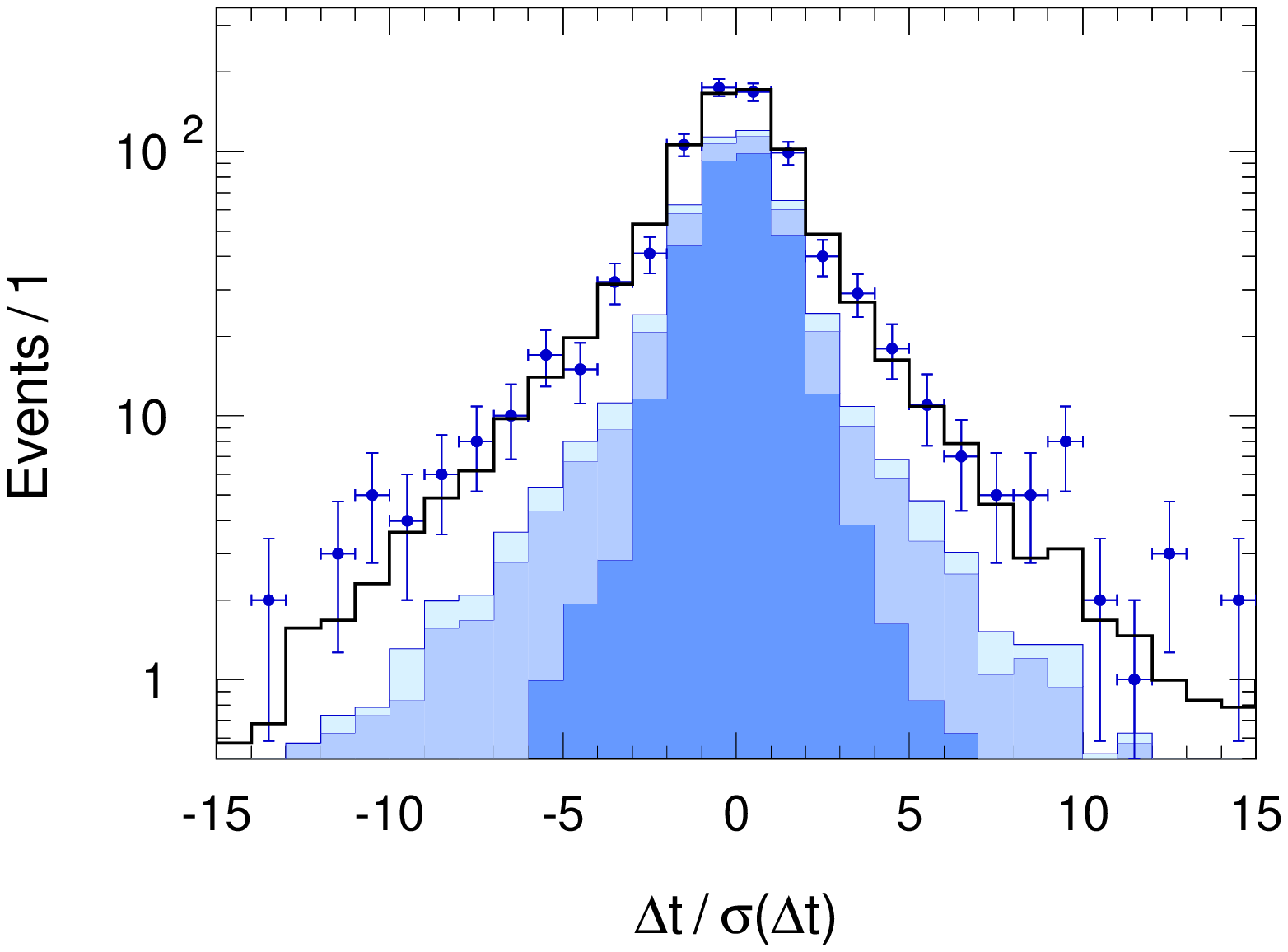}}
  \vspace{-0.1cm}
  \centerline{  \epsfxsize8.2cm\epsffile{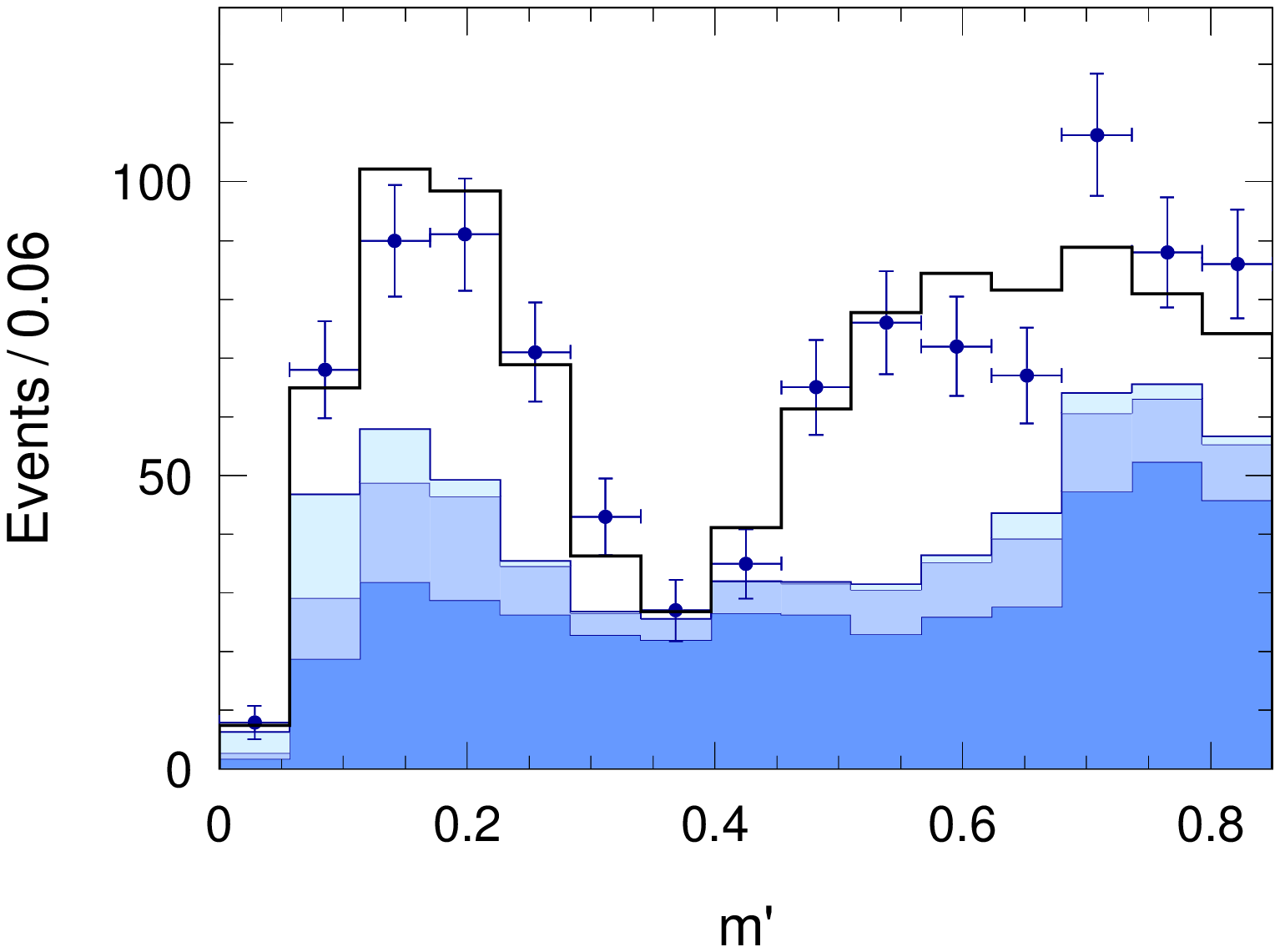}
                \epsfxsize8.2cm\epsffile{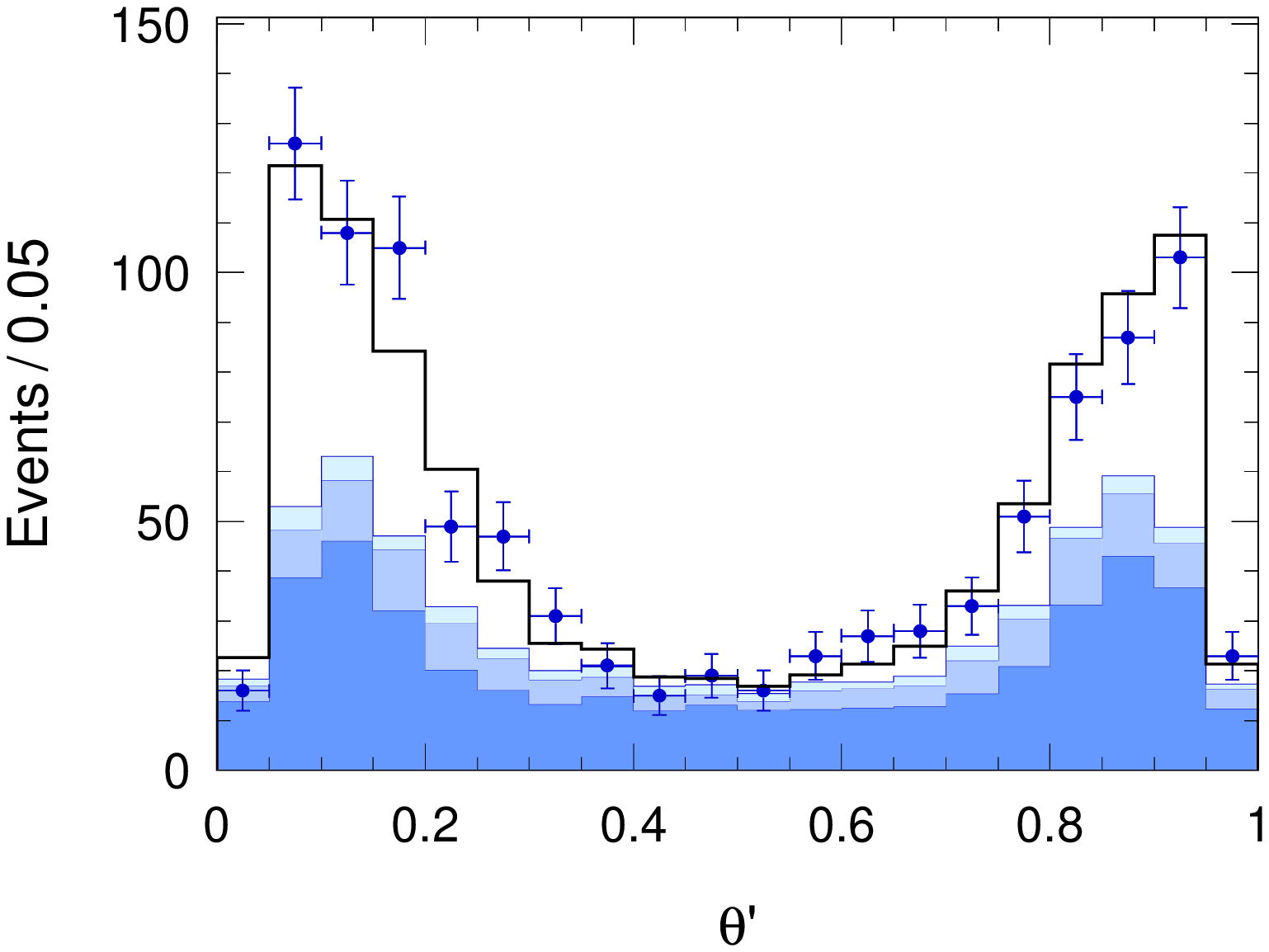}}
  \vspace{-0.5cm}
  \caption{\label{fig:projections} 
	Distributions of (clockwise from top left) $\deprime$, $\mes$, 
	NN output, $\dt/\sigma(\dt)$,  $\mprime$ and $\thetaprime$ for samples 
	enhanced in $\Btopipipi$ signal. The dots with error bars give 
	the on-resonance data. The solid histogram shows the
	projection of the fit result. The dark,
	medium and light shaded areas represent respectively the contribution
	from continuum events, the sum of continuum events 
	and the $B$-background expectation, and the sum of these and 
	the misreconstructed signal events. }
\end{figure}

The maximum-likelihood fit results in the $\Btopipipi$ event yield
$1184\pm58$, where the error is statistical only. For the $U$ and $I$
coefficients we find the results given together with their statistical
and systematic errors in Table~\ref{tab:results}. The corresponding 
correlation matrix is given in Table~\ref{tab:corrmat}. We have 
generated a sample of Monte Carlo experiments to determine the probability
density distributions of the fit parameters. Within the statistical 
uncertainties of this sample we find symmetric pulls and root-mean-squares
of order of unity for the $U$ and $I$ coefficients. This allows us to 
use the least-squares method to derive other quantities from these
(Section~\ref{sec:Physics}).

The signal is 
dominated by $\Bz\to\rho^\pm\pi^\mp$ decays. We observe an excess of
$\rho^0\piz$ events, which is in agreement with our previous upper
limit of $2.9 \times 10^{-6}$ at $90\%$ C.L.~\cite{rho0pi0paper}.
The result for the $\rho(1450)$ amplitude is in agreement with the findings 
in $\tau$ and $\epem$ decays~\cite{taueeref}. For the relative strong phase between 
the $\rho(770)$ and the $\rho(1450)$ amplitudes we find 
$(199^{\,+13}_{\,-17})^\circ$ (statistical error only), which is 
compatible with the result from $\tau$ and $\epem$ data.

Figure~\ref{fig:projections} shows distributions 
of $\deprime$, $\mes$, the NN output, $\dt/\sigma(\dt)$, where $\sigma(\dt)$
is the per-event error on $\dt$, as well as the Dalitz plot 
variables $\mprime$ and $\thetaprime$,
which are enhanced in signal content by cuts on the signal-to-continuum 
likelihood ratios of the other discriminating variables. 

As a validation of our treatment of the time dependence  we allow
$\tau_{\Bz}$ and $\dmd$ to vary in the fit. We find
$\tau_{\Bz} = (1.549\pm 0.080)\ps$ and $\dmd = (0.57\pm 0.11)\ps^{-1}$,
while the remaining free parameters are consistent with the nominal fit.
To validate the SCF modeling, we let the average SCF fractions per tagging
category free to vary in the fit and find results that are consistent
with the MC prediction. We also 
leave the $B$-background yields that normalize the $B$-background classes 
free to vary for those classes that have sufficiently different distributions 
of the fit variables to be distinguished from signal. The results are 
compatible with the model assumptions.

\section{INTERPRETATION OF THE RESULTS}
\label{sec:Physics}

The $U$ and $I$ coefficients are related to the 
quasi-two-body parameters, defined in Ref.~\cite{rhopipaper}\footnote
{
	For the cosine coefficients we adopt the convention:
	$C^+=[ \Gamma(B^0 \to \rho^+\pi^-) - \Gamma(\Bbar^0 \to \rho^+\pi^-) ] / [ \Gamma(B^0 \to \rho^+\pi^-) + \Gamma(\Bbar^0 \to \rho^+\pi^-) ]$ 
	and $C^-=[ \Gamma(B^0 \to \rho^-\pi^+) - \Gamma(\Bbar^0 \to \rho^-\pi^+) ] / [ \Gamma(B^0 \to \rho^-\pi^+) + \Gamma(\Bbar^0 \to \rho^-\pi^+) ]$.
}, 
as follows
\beqn
\label{eq:q2bparams}
	C^+ = \frac{ U^-_+ }{ U^+_+ }~, \hspace{0.6cm}
	C^- = \frac{ U^-_- }{ U^+_- }~, \hspace{0.6cm}
	S^+ = \frac{ 2 \, I_+ }{ U^+_+ }~, \hspace{0.6cm}
	S^- = \frac{ 2 \, I_- }{ U^+_- }~, \hspace{0.6cm}
	\Acp = \frac{ U^+_+ \, - U^+_- }{ U^+_+ \, + U^+_- }~,
\eeqn
and where $C=(C^++C^-)/2$, $\dC=(C^+-C^-)/2$, $S=(S^++S^-)/2$, 
and $\dS=(S^+-S^-)/2$. In contrast to our previous 
analysis~\cite{rhopipaper}, the definitions of Eq.~(\ref{eq:q2bparams})
explicitly account for the presence of interference effects, and are 
thus exact even for a $\rho$ with finite width, as long as the $U$ and 
$I$ coefficients are obtained with a Dalitz plot analysis. This treatment
leads to a dilution of the result and hence to slightly
increased statistical uncertainties compared to neglecting
the interference effects.

For the \CP-violation parameters, we obtain
\beqn
	\Acp    &=&      -0.088\pm 0.049 \pm{0.013}~, \\
	C   	&=& \ph{-}0.34 \pm 0.11 \pm 0.05~, \\
	S 	&=&      -0.10\pm 0.14\pm 0.04\,,
\eeqn
where the first errors given are statistical and the second systematic.
For the other parameters in the quasi-two-body description of the 
$\Bz(\Bzb) \to \rhopi$ decay-time dependence, we measure
\beqn
	\dC 	&=& 0.15\pm 0.11\pm 0.03~, \\
	\dS 	&=& 0.22\pm 0.15\pm 0.03~.
\eeqn
We find the linear correlation coefficients 
$c_{C,\dC}=0.20$ and $c_{S,\dS}=0.14$, while all 
other correlations are smaller. The systematic errors are dominated by the 
uncertainty on the \CP content of the \B-related backgrounds.
Other contributions are the signal description in the likelihood 
model (including the limit on non-resonant $\Btopipipi$ events), and
the fit bias uncertainty.

One can transform the experimentally convenient, namely
uncorrelated, direct \CP-violation parameters $\Crhopi$ and $\Acp$
into the physically more intuitive quantities $\Acppm$, $\Acpmp$,
defined by
\beqn
\label{eq:Adirpm}
    \Acppm &=& \frac{|\kappm|^2-1}{|\kappm|^2+1}
    \;=\; -\frac{\Acp+\Crhopi+\Acp\dCrhopi}{1+\dCrhopi+\Acp\Crhopi}
    ~,\\[0.2cm]\nonumber
\label{eq:Adirmp}
    \Acpmp &=& \frac{|\kapmp|^2-1}{|\kapmp|^2+1}
    \;=\; \frac{\Acp-\Crhopi-\Acp\dCrhopi}{1-\dCrhopi-\Acp\Crhopi}
    ~,
\eeqn
where
$\kappm = (q/p)(\Ampb/\Apm)$ and $\kapmp = (q/p)(\Apmb/\Amp)$,
so that $\Acppm$ ($\Acpmp$) involves only diagrams where the $\rho$
($\pi$) meson is emitted by the $W$ boson. We find
\begin{figure}[t]
  \centerline{  \epsfxsize7.7cm\epsffile{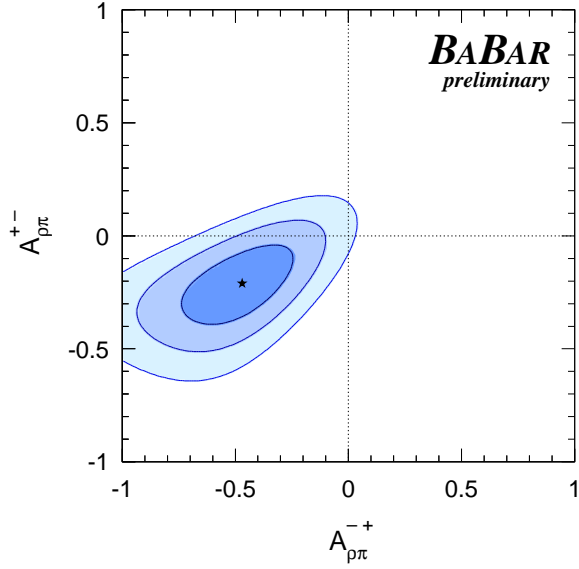}}
  \vspace{-0.1cm}
  \caption{\label{fig:apmamp} 
	Confidence level contours for the direct \CP asymmetries
	$\Acppm$ versus $\Acpmp$. The shaded areas represent 
	$1\sigma$, $2\sigma$ and $3\sigma$ contours, respectively. }
\end{figure}
\beqn
    \Acppm &=& -0.21\pm0.11\pm0.04~, \\
    \Acpmp &=& -0.47^{\,+0.14}_{\,-0.15}\pm0.06~,
\eeqn
with a correlation coefficient of 0.59 between $\Acppm$
and $\Acpmp$. The confidence level contours including systematic
errors are given in Fig.~\ref{fig:apmamp}. The significance, including 
systematics, for the observation of non-zero direct \CP violation is 
at the $2.9\sigma$ level.
\begin{figure}[t]
  \centerline{  \epsfxsize8.2cm\epsffile{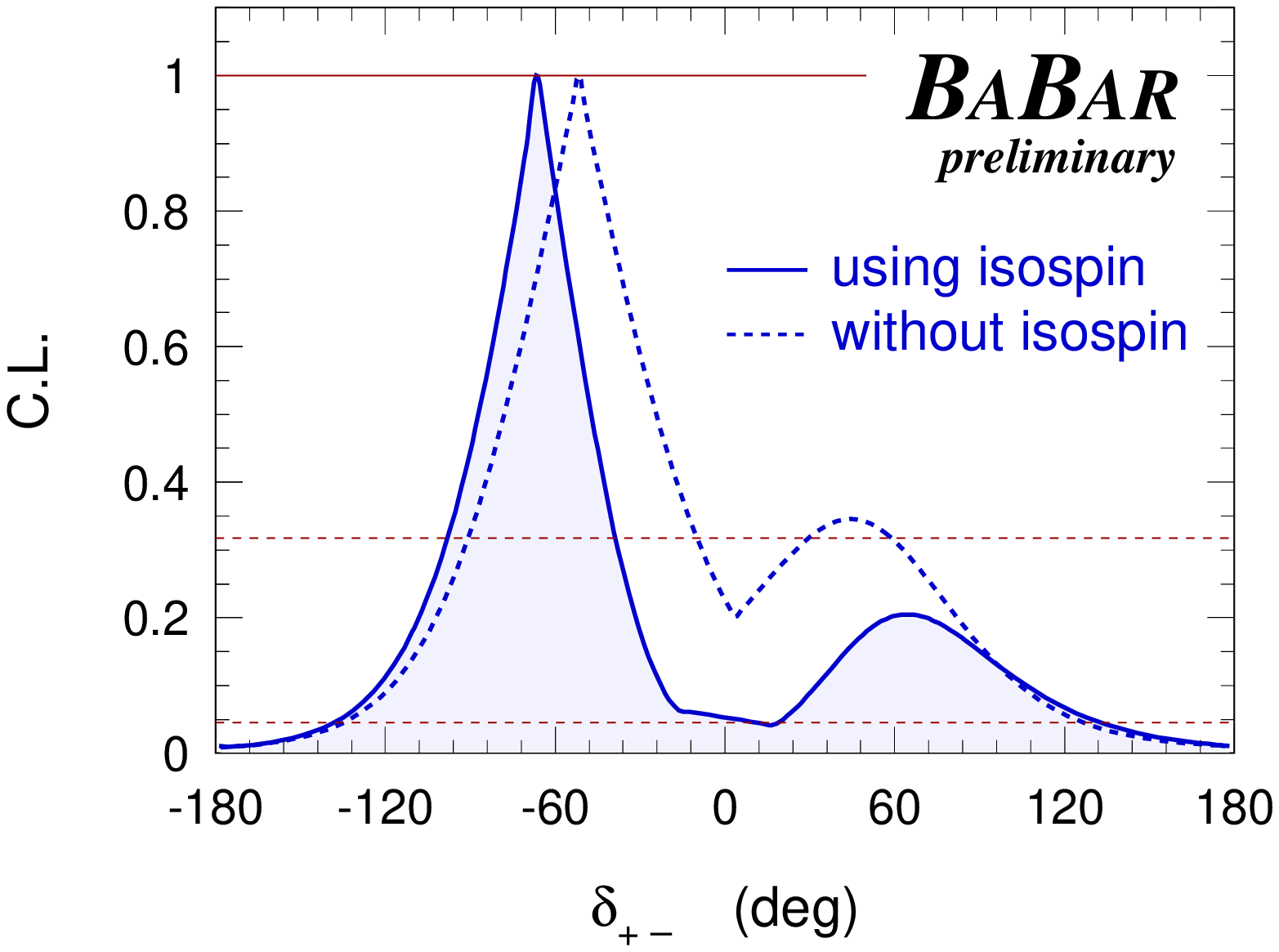}
   	        \epsfxsize8.2cm\epsffile{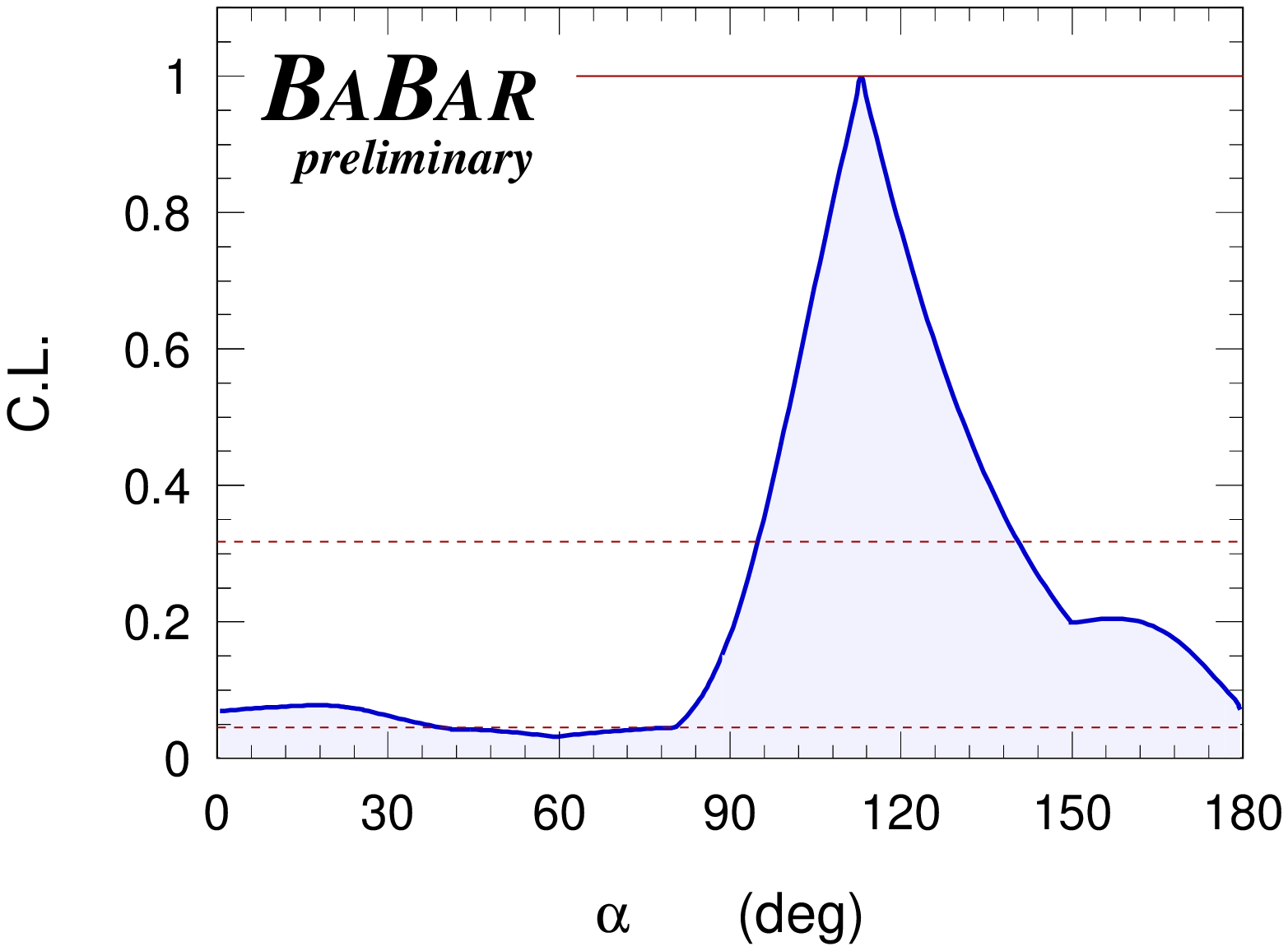}}
  \vspace{-0.4cm}
  \caption{\label{fig:deltaalpha} 
	Confidence level functions for $\delta_{+-}$ (left) and 
	$\alpha$ (right). Indicated 
	by the dashed horizontal lines are the C.L. values corresponding 
	to $1\sigma$ and $2\sigma$, respectively.}
\end{figure}

The measurement of the resonance interference terms allows us to 
determine the relative phase
\beq
\label{eq:deltapm}
\delta_{+-} = \arg\left( A^{+*}A^{-} \right)~,
\eeq
between the amplitudes of the 
decays $B^0\to\rho^-\pi^+$ and $B^0\to\rho^+\pi^-$. Through 
the definitions~(\ref{eq:firstObs})--(\ref{eq:lastObs}), we can derive 
a constraint on $\delta_{+-}$ from the measured $U$ and $I$ 
coefficients\footnote
{
	Using directly the relation 
	$\delta_{+-} = - \arctan\left[\left(U^{+,\mathrm{Im}}_{+-} + U^{-,\mathrm{Im}}_{+-}\right)/\left(U^{+,\mathrm{Re}}_{+-} + U^{-,\mathrm{Re}}_{+-}\right)\right]$ 
	to determine $\delta_{+-}$, leads only to a weak constraint,
	because it does not include all the available fit information.
} by 
performing a least-squares minimization with the six complex amplitudes
as free parameters. In this fit, one complex amplitude can be fixed due 
to an arbitrary global phase and the normalization condition $U_+^+=1$, 
leaving 10 real-valued unknowns. We obtain the confidence 
level (C.L.) function represented by the dashed line in the left 
hand plot of Fig.~\ref{fig:deltaalpha}. The function includes systematic 
errors.

This result does not require assumptions beyond the ones outlined in the 
introduction. The constraint can be improved with the use of strong 
isospin symmetry. The amplitudes 
$\Aij$ represent the sum of tree-level and penguin-type
amplitudes, which have different CKM factors: the tree-level $(\Tij)$
$\Bz\to\rho^\kappa\pi^{\kappab}$ transition amplitude\footnote
{
	We denote by $\kappab$ the charge conjugate of $\kappa$,
	where $\overline 0=0$. 
} 
is proportional to 
$V_{ud}V_{ub}^*$, while the corresponding penguin-type amplitude $(\Pij)$
comes with $V_{qd}V_{qb}^*$, where $q=u,c,t$. Using the unitarity of the 
CKM matrix one can reorganize the amplitudes and obtains~\cite{BaBarPhysBook}
\beqn
\label{eq:aijamps}
	\Aij 		&=& \Tij e^{-i\alpha} + \Pij ~, \nonumber\\
	\Abij 		&=& \Tji e^{+i\alpha} + \Pji ~,
\eeqn
where the magnitudes of the CKM factors have been absorbed in the 
$\Tij$, $\Pij$, $\Tji$ and $\Pji$. 
The Eqs.~(\ref{eq:aijamps}) represent 13 unknowns
of which two can be fixed due to an arbitrary
global phase and the normalization condition $U_+^+=1$. Using
strong isospin symmetry and neglecting isospin-breaking effects,
one can identify $P^{0}=-(P^{+}+P^{-})/2$, which reduces the 
number of unknowns to be determined by the fit to 9. This set of parameters 
provides the constraint on $\delta_{+-}$, given by the solid line on the 
left plot of Fig.~\ref{fig:deltaalpha}. The fit
returns a minimum $\chi^2$ of 7.4, which has a significance level of 0.39 
for 7 degrees of freedom. We find for the solution that
is favored by the fit
\beq
	\delta_{+-} \; = \; \left(-67^{\,+28}_{\,-31}\pm7\right)^\circ~,
\eeq
where the first errors are statistical and the second systematic.
Only a marginal constraint on $\delta_{+-}$ is obtained for ${\rm C.L.}<0.05$. 

Finally, following the same procedure, we can also
derive a constraint on $\alpha$ from the measured $U$ and $I$ 
coefficients. The resulting
C.L. function versus $\alpha$ is given in the right hand plot
of Fig.~\ref{fig:deltaalpha}. It includes systematic uncertainties.
Ignoring the mirror solution at $\alpha + 180^\circ$, we find
\beq
	\alpha \; = \; \left(113^{\,+27}_{\,-17}\pm6\right)^\circ~,
\eeq
where the systematic error is dominated by the uncertainties in the 
signal model. Only a marginal constraint on $\alpha$ is obtained for 
${\rm C.L.}<0.05$.

\section{SUMMARY}
\label{sec:Summary}

We have presented the preliminary measurement of 
\CP-violating asymmetries in $\Btopipipi$ decays dominated by 
the $\rho$ resonance. The results are obtained from a data sample 
of 213 millions $\FourS \to B\Bbar$ decays. We extend the previous 
quasi-two-body approach by using a time-dependent Dalitz plot 
analysis. From the measurement of the coefficients of 16 form 
factor bilinears we determine the three \CP-violating 
and two \CP-conserving quasi-two-body parameters, where we find a 
$2.9\sigma$ evidence of direct \CP violation. Taking advantage of 
the interference between the $\rho$ resonances in the Dalitz plot,
we derive constraints on the relative strong phase between 
$\Bz$ decays to $\rho^-\pip$ and $\rho^+\pim$, and on the angle
$\alpha$ of the Unitarity Triangle. These measurements are 
consistent with the expectation from the CKM fit~\cite{alphaSM}.

\section{ACKNOWLEDGMENTS}
\label{sec:Acknowledgments}


We are grateful for the 
extraordinary contributions of our \pep2\ colleagues in
achieving the excellent luminosity and machine conditions
that have made this work possible.
The success of this project also relies critically on the 
expertise and dedication of the computing organizations that 
support \babar.
The collaborating institutions wish to thank 
SLAC for its support and the kind hospitality extended to them. 
This work is supported by the
US Department of Energy
and National Science Foundation, the
Natural Sciences and Engineering Research Council (Canada),
Institute of High Energy Physics (China), the
Commissariat \`a l'Energie Atomique and
Institut National de Physique Nucl\'eaire et de Physique des Particules
(France), the
Bundesministerium f\"ur Bildung und Forschung and
Deutsche Forschungsgemeinschaft
(Germany), the
Istituto Nazionale di Fisica Nucleare (Italy),
the Foundation for Fundamental Research on Matter (The Netherlands),
the Research Council of Norway, the
Ministry of Science and Technology of the Russian Federation, and the
Particle Physics and Astronomy Research Council (United Kingdom). 
Individuals have received support from 
CONACyT (Mexico),
the A. P. Sloan Foundation, 
the Research Corporation,
and the Alexander von Humboldt Foundation.

\end{document}